\def\me{\hbox{\rm me}}
\documentclass[12pt,a4paper,onecolumn]{article}
\begin{document}
%\doublespacing
\title{Electromagnetic Waves in Variable Media}

\author{Ulrich Brosa\footnote{brosa-gmbh@t-online.de}\\
        Brosa GmbH, Am Br\"ucker Tor 4, D-35287 Am\"oneburg, Germany\\
        and Philipps-Universit\"at, Renthof 6, D-35032 Marburg}
\maketitle
\begin{abstract}
Two methods are explained to exactly solve Maxwell's equations
where permittivity, permeability and conductivity may vary in space.
In the constitutive relations, retardation is regarded. If the material
properties depend but on one coordinate, general solutions are derived. If
the properties depend on two coordinates, geometrically restricted solutions
are obtained. Applications to graded reflectors, especially
to dielectric mirrors, to filters, polarizers and to waveguides,
plain and cylindrical, are indicated. New foundations for the design of
optical instruments, which are centered around an axis, and for the design
of invisibility cloaks, plain and spherical, are proposed. The variability
of material properties makes possible effects which cannot happen
in constant media, e.g.\ stopping the flux of electromagnetic energy
without loss. As a consequence, spherical devices can be constructed
which bind electromagnetic waves.


\bigskip\noindent
(The final version appeared in Z.~Naturforsch.~{\bf 67a},~111-131~(2012).)
\end{abstract}
\section{Two steps towards reality}
\label{sec:two}
Let us trace the way Schr\"odinger walked to find the
Schr\"odinger equation. He thought that the propagation of light,
if it is construed as propagation of particles, is best
described by the eikonal equation:
\begin{equation}
(\nabla s({\bf r}))^2=n^2({\bf r})\ .                         \label{eik}
\end{equation}
Surfaces of equal eikonal $s({\bf r})$ are perpendicular to the light rays
everywhere in the space described by the vector of location {\bf r}.
$n({\bf r})$ is the index of refraction. Schr\"odinger
compared this with the Helmholtz equation
\begin{equation}
\nabla^2 \psi({\bf r})+{n^2({\bf r})\over c^2}\omega^2\psi({\bf r})=0 \label{hel}
\end{equation}
which he considered as the best description of waves.
$\omega$ denotes the frequency of that wave
and $c$ is the velocity of light. The meaning of $\psi$ is not known.
Next Schr\"odinger remembered that there is an eikonal equation
for massive particles, too, the Hamilton-Jacobi equation
\begin{equation}
(\nabla S({\bf r}))^2=2m(E-V({\bf r}))\ .                      \label{ham}
\end{equation}
The mechanical eikonal $S({\bf r})$ has a similar meaning as in
ray optics, but its dimension is different. Hence Schr\"odinger
deduced from a comparison of (\ref{eik}) and (\ref{ham}) a mechanical
index of refraction
\begin{equation}
n({\bf r})=\sqrt{{c^2\over \hbar^2\omega} 2m(E-V({\bf r}))}\ . \label{ref}
\end{equation}
The factor in front of $2m(E-V({\bf r}))$ is an adjustable constant
to get dimensions right. That it is related to Planck's constant $\hbar$,
Schr\"odinger realized when he solved the first problems. However,
when the guess (\ref{ref}) is used in the Helmholtz equation (\ref{hel}),
an equation for the wavy propagation of massive particles is established,
the Schr\"odinger equation:
\begin{equation}
\nabla^2 \Psi({\bf r})+{2m\over\hbar^2}(E-V({\bf r}))\Psi({\bf r})=0\ .
                                                                \label{sch}
\end{equation}
Here also the meaning of $\Psi$ is not clear.

The problem with this type of approach is the Helmholtz equation (\ref{hel}).
The wavy propagation of light is reigned by Maxwell's equations.
In the analytic solution of these equations, the Helmholtz equation occurs
as a mathematical auxiliary \cite{Bro10}, but this is only true when
material properties as permittivity $\varepsilon$ and permeability $\mu$
are constant. What is the replacement of the Helmholtz equation
if these properties and thus the index of refraction
\begin{equation}
n({\bf r})=
\sqrt{{\epsilon_0\mu_0\over\varepsilon({\bf r})\mu({\bf r})}}  \label{ind}
\end{equation}
vary in space? This question will be answered in Sections \ref{sec:triple}
and \ref{sec:alternative}. The modifications
will turn out so severe that the Helmholtz equation can be considered
only in rare cases as an approximation. It is not even possible
to formulate the true equations using the index of refraction only.
Permittivity and permeability enter individually.

The true companion of Maxwell's equations is Dirac's equation.
Maxwell's is for vectors, Dirac's for spinors. Yet both systems
carry similar information, namely equations for divergences
and curls related to time derivatives. Dirac's equation is
a linear system of partial differential equations with
variable coefficients, the electrodynamic potentials.
In Maxwell's equations, variable coefficients appear when
permittivity, permeability and conductivity depend on location.
Dirac's equation can be solved analytically
if the coefficients vary just one-dimensionally or
if they vary central-symmetrically. The analog for Maxwell's
equations, and more, is the main result of this article,
see Sections \ref{sec:one} and \ref{sec:central}.

The impact of this article should be even larger on
practical problems. In modern times, people fabricate graded materials
or so-called metamaterials within which permittivity and permeability
vary in space almost arbitrarily. Therefore analytic solutions that
predict effects of such variations will be useful.

Yet usefulness for practicians coerces the consideration
of dissipation and dispersion. Most materials have finite conductivity.
Ohmic currents must be included in the theory. Moreover inertia and
friction within the materials modify permittivity $\varepsilon$,
permeability $\mu$ and conductivity $\sigma$. The simple constants
must be upgraded, in a minimum approach to reality, to response functions
which vary in space and describe retardation:
\begin{eqnarray}
{\bf D}({\bf r},t)=\int_0^t \varepsilon({\bf r},t-\tau)
     {\bf E}({\bf r},\tau)\hbox{d}\tau\ ,                    \label{resp1}\\
{\bf B}({\bf r},t)=\int_0^t \mu({\bf r},t-\tau)
     {\bf H}({\bf r},\tau)\hbox{d}\tau\ ,                    \label{resp2}\\
{\bf j}({\bf r},t)=\int_0^t \sigma({\bf r},t-\tau)
     {\bf E}({\bf r},\tau)\hbox{d}\tau\ .                    \label{resp3}
\end{eqnarray}
${\bf D}({\bf r},t)$, ${\bf E}({\bf r},t)$, ${\bf B}({\bf r},t)$,
${\bf H}({\bf r},t)$ and ${\bf j}({\bf r},t)$ denote dielectric
displacement, electric force field, magnetic force field, magnetic
field strength and electric current density, respectively.
They all are vector fields depending on space ${\bf r}$ and time $t$.

With the constitutive relations (\ref{resp1}-\ref{resp3})
the evolution of the electrodynamic field is completely conceived
by Maxwell's equations
\begin{eqnarray}
\nabla\times{\bf E}({\bf r},t)=-\partial_t{\bf B}({\bf r},t)\ , \label{max1}\\
\nabla{\bf B}({\bf r},t)=0\ ,                                   \label{max2}\\
\nabla\times{\bf H}({\bf r},t)=\partial_t{\bf D}({\bf r},t)
+{\bf j}({\bf r},t)\ ,                                          \label{max3}\\
\nabla{\bf D}({\bf r},t)=\rho({\bf r},t)\ ,                     \label{max4}
\end{eqnarray}
written in an unfamiliar sequence for reasons
that will become clear in Section \ref{sec:reshaping}.

So these are the two steps to reality: First, solutions of Maxwell's
equations shall be found with material properties that vary in space.
Second, retardation shall be taken into account.

Yet generality will be restricted in two ways: First, only the homogeneous
problem will be tackled. For example, externally driven currents
will be omitted. The reason to keep nevertheless the charge density
$\rho({\bf r},t)$ and the current density ${\bf j}({\bf r},t)$
in Maxwell's equations is to admit Ohmic currents.
The exclusion of nonhomogeneities is not a serious limitation
as there are standard procedures to construct the solutions
of nonhomogeneous equations from the solutions of the homogeneous system.

By contrast, the second lack cannot be cured and
can be justified only by the desire to produce exact solutions
of Maxwell's equations: All material properties will be restricted
to depend on one or two spatial coordinates only. To be specific, introduce
coordinates $\xi,\eta,\zeta$ to describe the vector of position {\bf r}.
They may be the Cartesian coordinates $x,y,z$, but generally
these greek letters are meant to describe curvilinear
yet orthogonal coordinates, for example spherical or cylindrical ones.
The normalized basis vectors shall be denoted as ${\bf e}_\xi$,
${\bf e}_\eta$, ${\bf e}_\zeta$ and the line element $ds$ be given as
\begin{equation}
(ds)^2=g_{\xi\xi}(d\xi)^2+g_{\eta\eta}(d\eta)^2+g_{\zeta\zeta}(d\zeta)^2
                                                                \label{line}
\end{equation}
with elements $g_{\xi\xi}$, $g_{\eta\eta}$, $g_{\zeta\zeta}$
of the metric tensor \cite{Moo61}. The response functions are supposed
to depend on $\zeta$ only
\begin{equation}
\varepsilon(\zeta,t),\quad\mu(\zeta,t),\quad\sigma(\zeta,t),   \label{respo1}
\end{equation}
see Section \ref{sec:triple}, or only on $\eta$ and $\zeta$
\begin{equation}
\varepsilon(\eta,\zeta,t),\quad\mu(\eta,\zeta,t),\quad\sigma(\eta,\zeta,t),
                                                                \label{respo2}
\end{equation}
see Section \ref{sec:alternative}.
In the second case (\ref{respo2}), which appears to be more general,
we will have to impose restrictions upon the solutions of Maxwell's
equations. Nevertheless the realm of exactly solvable problems will
be extended immensely.

This is the plan of this article: In Section \ref{sec:reshaping} Maxwell's
and the constitutive equations will be rewritten to facilitate a simple
description of retardation. In Sections \ref{sec:triple} and
\ref{sec:alternative} the main results will be produced and proven, namely
two theorems of representation. They reduce the eight mingled Maxwellian
equations to two uncoupled partial differential equations each for one
unknown only.
Applications of these theorems are sketched in the Sections \ref{sec:one},
\ref{sec:graded}, \ref{sec:stopping}, \ref{sec:mirror},
\ref{sec:central}, \ref{sec:bound}, \ref{sec:plane} and \ref{sec:axis}.
Finally in Section \ref{sec:retrospect} attempts are made to do justice
to precursors of the ideas and the results presented here.

\section{Reshaping Maxwell's equations}
\label{sec:reshaping}
The first two equations (\ref{max1}-\ref{max2}) are ideally simple.
The aim is to rewrite the last two equations (\ref{max3}-\ref{max4})
until they take the same shape as the first two. The clue is to introduce
the {\it complete displacement\/}
\begin{equation}
{\bf C}({\bf r},t)={\bf D}({\bf r},t)+
\int_0^t{\bf j}({\bf r},\tau)\hbox{d}\tau\ .                   \label{comp}
\end{equation}
This takes (\ref{max3}) to
\begin{equation}
\nabla\times{\bf H}({\bf r},t)=\partial_t{\bf C}({\bf r},t) \label{max3s}
\end{equation}
having up to a sign the same structure as (\ref{max1}). Similarly
(\ref{max4}) appears as
\begin{equation}
\nabla{\bf C}({\bf r},t)=0                                 \label{max4s}
\end{equation}
having the exactly same structure as (\ref{max2}). Because of the
continuitity equation
\begin{equation}
\rho({\bf r},t)=-\nabla\int_0^t{\bf j}({\bf r},\tau)\hbox{d}\tau+
\rho({\bf r},0)                                            \label{cont}
\end{equation}
equation (\ref{max4s}) is true if there is no initial bunching of charges
$\rho({\bf r},0)=0$.
The effects of an initial bunching of charges can be covered by
a scalar potential in a manner explained in \cite[Sec.2]{Bro10}. It is
not extraordinary enough to be treated here.

The reshaped Maxwell equations are now (\ref{max1}-\ref{max2})
and (\ref{max3s}-\ref{max4s}).
To close the system, we have to combine from
(\ref{resp1}) and (\ref{resp3}) the constitutive equation for the complete
displacement. It is
\begin{eqnarray}
{\bf C}({\bf r},t)=\int_0^t\epsilon({\bf r},t-\tau)
     {\bf E}({\bf r},\tau)\hbox{d}\tau                      \label{resp4}\\
\epsilon({\bf r},t)=\varepsilon({\bf r},t)+
     \int_0^t\sigma({\bf r},\tau)\hbox{d}\tau\ \             \label{cresp}
\end{eqnarray}
with the {\it complete permittivity\/} $\epsilon({\bf r},t)$.

The reshaped Maxwell equations (\ref{max1}-\ref{max2})
and (\ref{max3s}-\ref{max4s}) together with the constitutive equations
(\ref{resp2}) and (\ref{resp4}) form a closed system, but it is
a system of integro-differential equations. Performing Laplace transforms
\begin{equation}
f_\omega({\bf r})=\int_0^\infty f({\bf r},t)
e^{i\omega t}\hbox{d}t                                   \label{lap}
\end{equation}
where $f({\bf r},t)$ may denote any component of
the vector fields or any response function,
one gets rid of the integrals. The convolution theorem converts
the integrals in the constitutive relations
(\ref{resp2}) and (\ref{resp4}) to products:
\begin{eqnarray}
{\bf B}_\omega({\bf r})=\mu_\omega({\bf r})\,{\bf H}_\omega({\bf r})\ ,
                                                              \label{con1}\\
{\bf C}_\omega({\bf r})=\epsilon_\omega({\bf r})\,{\bf E}_\omega({\bf r})\ .
                                                              \label{con2}
\end{eqnarray}
The Laplace transform of the complete permittivity follows from
(\ref{cresp})
\begin{equation}
\epsilon_\omega({\bf r})=\epsilon_\omega({\bf r})
                       +i\sigma_\omega({\bf r})/\omega\ .     \label{crespi}
\end{equation}
It is just a linear combination of the Laplace transforms of the ordinary
permittivity and the conductivity.

Also one gets rid of the derivatives with respect to time $t$:
\begin{equation}
\int_0^\infty \partial_t f({\bf r},t) e^{i\omega t}\hbox{d}t
=-i\omega f_\omega({\bf r})-f({\bf r},0)\ .                   \label{lapd}
\end{equation}
The second term on the right-hand side is a valuable peculiarity
of the Laplace transform as it facilitates straightforward solutions
of initial-value problems.

The Maxwell equations (\ref{max1}-\ref{max2}) and (\ref{max3s}-\ref{max4s})
are transformed to
\begin{eqnarray}
\nabla\times{\bf E}_\omega({\bf r})=
                      i\omega\,{\bf B}_\omega({\bf r})\ ,     \label{max1a}\\
\nabla{\bf B}_\omega({\bf r})=0\ ,                            \label{max2a}\\
\nabla\times{\bf H}_\omega({\bf r})=
                     -i\omega\,{\bf C}_\omega({\bf r})\ ,     \label{max3a}\\
\nabla{\bf C}_\omega({\bf r})=0\ .                            \label{max4a}
\end{eqnarray}
Here two other nonhomogeneities were omitted,
viz.\ ${\bf B}({\bf r},0)$ and $-{\bf C}({\bf r},0)$
on the right-hand sides of (\ref{max1a}) and (\ref{max3a})
which arise from the Laplace transforms of
$-\partial_t{\bf B}({\bf r},t)$ and
$\partial_t{\bf C}({\bf r},t)$, respectively, according to
equation (\ref{lapd}).

Most people would consider the reshaped Maxwell equations
(\ref{max1a}-\ref{max4a}) as obtained from the orginal Maxwell equations
just by separation of $\exp(-i\omega t)$. Yet this point of view
hides the origin of the permittivity $\epsilon_\omega({\bf r})$ and
permeability $\mu_\omega({\bf r})$ depending on
{\it frequency\/} and it aggravates the solution of initial-value problems,
i.e.\ it impedes a rational theory of pulses. To do this, one has, first,
to solve the reshaped Maxwell equations (\ref{max1a}-\ref{max4a}) with
the reshaped constitutive relations (\ref{con1}-\ref{con2}), second, to
introduce the initial-values as nonhomogeneities in (\ref{max1a}) and
(\ref{max3a}) and to solve the nonhomogeneous system and, third, to calculate
the pulses from the inverse Laplace transform:
\begin{equation}
f({\bf r},t)={1\over 2\pi}\int_{-\infty+ir}^{\infty+ir}f_\omega({\bf r})
e^{-i\omega t}\hbox{d}\omega\ .                                 \label{lapi}
\end{equation}
$r$ denotes a real number big enough such that all locations of sigularities
of $f_\omega({\bf r})$ in the complex plane of $\omega$ have smaller
real parts.

The customary variable of Laplace transforms is $p=i\omega$.
The author introduced $\omega$ instead in order to pacify conservative
readers. It they want to believe that the equations
(\ref{max1a}-\ref{max4a}) are just the ordinary Maxwell equations
with $\exp(-i\omega t)$ separated off, they can do so.
The calculations to be presented right now, however, do not depend
on this point of view. So let us abbreviate:
\begin{eqnarray}
{\bf C}={\bf C}_\omega({\bf r}),\
{\bf E}={\bf E}_\omega({\bf r}),\
\epsilon=\epsilon_\omega({\bf r}),\ \nonumber\\
{\bf B}={\bf B}_\omega({\bf r}),\
{\bf H}={\bf H}_\omega({\bf r}),\
\mu=\mu_\omega({\bf r}).                                     \label{abbr}
\end{eqnarray}

\section{Triple curl again}
\label{sec:triple}
The aim is to reduce all Maxwell equations (\ref{max1a}-\ref{max4a})
to one partial differential equation for one scalar auxiliary, the
representative $b=b_\omega({\bf r})$. The approach is
the same as in \cite[Sec.2]{Bro10} looking for certain equations
with triple curl.

The ansatz
\begin{eqnarray}
{\bf B}=-\nabla\times\nabla\times{\bf v}b                    \label{ans1}\\
{\bf E}=-i\omega\nabla\times{\bf v}b                         \label{ans2}
\end{eqnarray}
solves two Maxwell equations immediately, viz.\ (\ref{max1a}) and
(\ref{max2a}). The vector field ${\bf v}$, the {\it carrier\/},
shall be chosen such that the remaining two equations (\ref{max3a})
and (\ref{max4a}), too, can be solved. Such a choice will be possible
if the response functions depend only on one spatial variable,
say $\zeta$, as declared in (\ref{respo1}), and thus
\begin{equation}
\epsilon=\epsilon_\omega(\zeta),\qquad\mu=\mu_\omega(\zeta).\label{respoo}
\end{equation}

Inserting the constitutive relations (\ref{con1}-\ref{con2})
into the ansatz (\ref{ans1}-\ref{ans2}) gives
\begin{eqnarray}
{\bf H}=-{1\over\mu}\nabla\times\nabla\times{\bf v}b\ ,      \label{ans1a}\\
{\bf C}=-i\omega\nabla\times{\bf v}\epsilon b\ .             \label{ans2a}
\end{eqnarray}
In the last equation the permittivity $\epsilon$ was drawn under the
curl though it is not constant. This can be justified
if the carrier ${\bf v}$ is chosen to point into the direction
of the basis vector ${\bf e}_\zeta$.
\begin{equation}
{\bf v}=|{\bf v}|{\bf e}_\zeta .                             \label{vco1}
\end{equation}
Then, because of (\ref{respoo}), the carrier points into
the same direction as the gradient of $\epsilon$
\begin{equation}
\nabla\epsilon=
{{\bf e}_\zeta\over\sqrt{g_{\zeta\zeta}}}\
{\partial\epsilon\over\partial\zeta}=
{{\bf v}\over|{\bf v}|\sqrt{g_{\zeta\zeta}}}\
{\hbox{d}\epsilon\over\hbox{d}\zeta}\ .                       \label{com}
\end{equation}
Consequently in the identity
\begin{equation}
\epsilon\nabla\times{\bf v}b =
\nabla\times{\bf v}\epsilon b+{\bf v}b\times\nabla\epsilon   \label{com1}
\end{equation}
the last term is zero and thus equation (\ref{ans2a}) proven.

Because of (\ref{ans2a}), Maxwell's equation (\ref{max4a}) is automatically
fulfilled, too. So we just have to care for (\ref{max3a}). Insertion of
(\ref{ans1a}) and (\ref{ans2a}) produces an equation of triple curl
\begin{equation}
\nabla\times(-{1\over\mu}\nabla\times\nabla\times{\bf v}b
+{\bf v}\epsilon\omega^2 b)=0\ .                              \label{tricur}
\end{equation}
The second term behind the leading curl is proportional to the carrier.
All that remains to be done in order to obtain the desired
scalar equation is to show that the first term is a gradient plus
a term which also aligns with the carrier.
First, we replace the double curl with the Laplacian $\nabla^2$:
\begin{equation}
-{1\over\mu}\nabla\times\nabla\times{\bf v}b=
{1\over\mu}(\nabla^2{\bf v}b-\nabla(\nabla{\bf v}b))\ .         \label{tricur1}
\end{equation}
Second, we commute the carrier {\bf v} with the Laplacian and require
that the commutation does not produce more than a gradient.
This can be done only if the carrier varies at most linearly
\begin{equation}
{\bf v}={\bf v}_0+v_1{\bf r}\quad\Leftrightarrow\quad
\nabla^2{\bf v}b={\bf v}\nabla^2 b+2v_1\nabla b\ ,            \label{tricur2}
\end{equation}
${\bf v}_0$ denoting a constant vector and $v_1$ a constant number.
The equivalence is valid only if the dependence of $b$ on
the coordinates is not restricted. For details of the proof
see \cite{Gro93} or \cite{Bro86}.
If $\mu$ were constant, we had completed the task. Then the second terms
on the right-hand sides of (\ref{tricur1}) and (\ref{tricur2}) are gradients
which the leading curl in (\ref{tricur}) discards. When $\mu$ varies,
we must effect a third transformation
\begin{equation}
{1\over\mu}(-\nabla(\nabla{\bf v}b)+2v_1\nabla b)=
\bigg(\nabla{1\over\mu}\bigg)(\nabla{\bf v}b -2v_1 b)-
\nabla{1\over\mu}(\nabla{\bf v}b -2v_1 b)\ .                    \label{tricur3}
\end{equation}
Here, at last, the second term on the right-hand side is a gradient,
while the first aligns with the carrier because of (\ref{respoo}).
Using again (\ref{vco1}) we find
\begin{equation}
\nabla{1\over\mu}=
{{\bf v}\over|{\bf v}|\sqrt{g_{\zeta\zeta}}}\
{\hbox{d}\over\hbox{d}\zeta}{1\over\mu}                     \label{tricur4}
\end{equation}
similar to (\ref{com}).
Collecting (\ref{tricur1}), (\ref{tricur2}) and (\ref{tricur3}) we derive
from (\ref{tricur}) the scalar equation
\begin{equation}
\nabla^2 b
+ {\mu\over|{\bf v}|\sqrt{g_{\zeta\zeta}}}
\bigg({\hbox{d}\over\hbox{d}\zeta}{1\over\mu}\bigg)(\nabla{\bf v}b -2v_1 b)
+ \epsilon\mu\omega^2 b = 0\ .                                 \label{repb}
\end{equation}

To find formulae for the other polarization, we stipulate the ansatz
\begin{eqnarray}
{\bf C}=\nabla\times\nabla\times{\bf v}a                    \label{ans1b}\\
{\bf H}=-i\omega\nabla\times{\bf v}a                        \label{ans2b}
\end{eqnarray}
with the representative $a=a_\omega({\bf r})$,
which automatically satisfies all Maxwell equations except (\ref{max1a}).
The demand to have also this one solved produces a triple-curl equation
similar to (\ref{tricur}). $a$ replaces $b$, $\epsilon$ and $\mu$ are
interchanged. Performing the same transformations as before, we arrive
at a scalar equation similar to (\ref{repb}).

Thus we finished the proof of the

\medskip
\noindent{\bf Three-Dimensional Representation Theorem.}
{\sl Solutions of Maxwell's equations (\ref{max1a}-\ref{max4a})
are provided by the representations
\begin{eqnarray}
{\bf E}={1\over\epsilon}\nabla\times\nabla\times{\bf v}a-
          i\omega\nabla\times{\bf v}b\phantom{-}               \label{sol1}\\
{\bf H}=-i\omega\nabla\times{\bf v}a-
         {1\over\mu}\nabla\times\nabla\times{\bf v}b           \label{sol2}
\end{eqnarray}
if the representatives $a$ and $b$ obey the scalar partial differential
equations
\begin{eqnarray}
\nabla^2 a -
{1\over{\bf v}^2}({\bf v}\nabla\log{\epsilon})
(\nabla{\bf v}a -2v_1 a) +
\epsilon\mu\omega^2 a=0                                        \label{rep1a}\\
\nabla^2 b -
{1\over{\bf v}^2}({\bf v}\nabla\log{\mu})
(\nabla{\bf v}b -2v_1 b) +
\epsilon\mu\omega^2 b=0                                        \label{rep1b}
\end{eqnarray}
and the carrier is chosen such that
\begin{equation}
{\bf v}={\bf v}_0+v_1{\bf r}\quad\hbox{and}\quad\nabla\epsilon\propto{\bf v}
\quad\hbox{and}\quad\nabla\mu\propto{\bf v}                    \label{dir}
\end{equation}}
${\bf v}_0$ being a constant vector and $v_1$ a constant number.
\medskip

In most optical instruments, different media meet. Often a
graded medium is surrounded by air. Usually material properties jump
discontinously where the media touch. Therefore boundary
conditions for the representatives are necessary.

\noindent{\bf Corollary on boundary-value conditions.} {\sl Let $S$ denote
the surface where different media meet, {\bf n} the normal
on this surface and $\partial/\partial n$ the differentiation along
this normal. If ${\bf v}\propto{\bf n}$, the representatives $a$ and $b$
must satisfy
\begin{eqnarray}
a\,|_{S-}=a\,|_{S+}\ ,\qquad
{1\over\epsilon}{\partial|{\bf v}|a\over\partial n}\,\bigg|_{S-}=
{1\over\epsilon}{\partial|{\bf v}|a\over\partial n}\,\bigg|_{S+}\ ,\label{boua0}\\
b\,|_{S-}=b\,|_{S+}\ ,\qquad
{1\over\mu}{\partial|{\bf v}|b\over\partial n}\,\bigg|_{S-}=
{1\over\mu}{\partial|{\bf v}|b\over\partial n}\,\bigg|_{S+}\ .     \label{boub0}
\end{eqnarray}
The symbols $S-$ and $S+$ indicate that the values of the functions
and their derivatives are to be calculated via an approach on the
one side of $S$, say, the low side $S-$, or on the other side, say, the high
side $S+$.}

Proof: It follows from the structure of the Maxwell equations
(\ref{max1a}) and (\ref{max3a})
that the tangential components of the magnetic field
strength {\bf H} and the electric force field {\bf E} don't jump on
transition through $S$. Evaluating these facts in the representation formulae
(\ref{sol1}-\ref{sol2}) produces the proof of the corollary. Details of the
calculational procedure are similar as in \cite[Sec.3]{Bro10}.
Q.E.D.

In optics it is difficult to observe the electromagnetic field directly.
Instead one measures the flux of energy which can be calculated
as the Pointing vector
${\bf S}({\bf r},t)={\bf E}({\bf r},t)\times{\bf H}({\bf r},t)$.
This is a general formula for physical fields ${\bf E}({\bf r},t)$ and
${\bf H}({\bf r},t)$. Before we can use it for the mathematical fields
handled here, we must calculate the dependence on time from (\ref{lapi})
and extract the real parts $\Re$. However, if the dependence on time can be
described by the factor $\exp({-i\omega t})$ with real frequency $\omega$,
we may apply the

\noindent{\bf Corollary on the energy flux.}
{\sl The time-averaged Pointing vector $\bar{\bf S}$ can be calculated
from the representatives $a$ and $b$ according to
\begin{eqnarray}
\bar{\bf S}={1\over 2}\Re({\bf E}\times{\bf H}^*)
= \Re\bigg({i\omega\over 2\epsilon}
(\nabla\times\nabla\times{\bf v}a)\times(\nabla\times{\bf v}a^*)\bigg)
                                             \hbox to 32pt{\hfil} \nonumber\\
+ \Re\bigg({i\omega\over 2\mu}
(\nabla\times{\bf v}b)\times(\nabla\times\nabla\times{\bf v}b^*)\bigg)
                                             \hbox to 32pt{\hfil} \nonumber\\
+ \Re\bigg({\omega^2\over 2}
(\nabla\times{\bf v}b)\times(\nabla\times{\bf v}a^*)\bigg)
                                             \hbox to 56pt{\hfil} \nonumber\\
- \Re\bigg({1\over 2\epsilon\mu}
(\nabla\times\nabla\times{\bf v}a)\times
(\nabla\times\nabla\times{\bf v}b^*)\bigg)                        \label{poi1}
\end{eqnarray}
the asterisk $^*$ denoting complex conjugation.}

The proof follows immediately from the representation formulae
(\ref{sol1}-\ref{sol2}).

When there is only one polarization, i.e. either $a=0$ or $b=0$,
the mixed terms in the third and the forth lines of (\ref{poi1})
do not apply. In the first and second lines,
reader's attention shouldn't miss the inconspicuous imaginary units $i$
and the factors $1/\epsilon$ as well as $1/\mu$. The former are
indispensable for a weird stop of energy flux in dielectric materials,
whereas the latter may alter the type of the flux considerably
when they are non-constant.

Due to the condition (\ref{dir}), two limiting cases stand out, namely
when the material properties vary one-dimensionally, see Section \ref{sec:one},
or central-symmetrically, see Section \ref{sec:central}.

\subsection{One-dimensional variations of material properties}
\label{sec:one}
$z$ be the name of the coordinate along which permittivity,
permeability and conductivity are allowed to vary.
It is the same $z$ which is costumary in the cartesian system and all
cylindrical coordinate systems. In the equations of Sections of
\ref{sec:two} and \ref{sec:triple} we got to set $\zeta=z$.
The unit vector along $z$ is chosen as carrier, i.e.\ ${\bf v}={\bf e}_z$.
According to condition (\ref{dir}) we have ${\bf v}_0={\bf e}_z$ and $v_1=0$.
The differential equations (\ref{rep1a}-\ref{rep1b}) become
\begin{eqnarray}
\nabla^2 a
-{\hbox{d}\log{\epsilon}\over\hbox{d}z}{\partial a\over\partial z}
+\epsilon\mu\omega^2 a=0\ ,                                  \label{repa1a}\\
\nabla^2 b
-{\hbox{d}\log{\mu}\over\hbox{d}z}{\partial b\over\partial z}
+\epsilon\mu\omega^2 b=0\ .                                  \label{repa1b}
\end{eqnarray}
These differential equations are valid in all coordinate systems
which incorporate a cartesian direction.

The most elementary example is the cartesian system $x,y,z$.
(\ref{repa1a}) appears as
\begin{equation}
{\partial^2 a\over \partial x^2}+
{\partial^2 a\over \partial y^2}+
{\partial^2 a\over \partial z^2}-
{\hbox{d}\log{\epsilon}\over\hbox{d}z}{\partial a\over\partial z}
+\epsilon\mu\omega^2 a=0 ,                                 \label{rep2a}
\end{equation}
which is separated by the ansatz
\begin{equation}
a=X\,Y\,Z_a\quad\hbox{with}\quad
X=X(x),\ Y=Y(y),\  Z_a=Z_{a\omega}(z)         \label{sepcar}
\end{equation}
to yield three ordinary differential equations
\begin{eqnarray}
{\hbox{d}^2 X\over\hbox{d}x^2}+k_x^2 X=0                    \label{car1}\\
{\hbox{d}^2 Y\over\hbox{d}y^2}+k_y^2 Y=0                    \label{car2}\\
{\hbox{d}^2 Z_a\over\hbox{d}z^2}
-{\hbox{d}\log{\epsilon}\over\hbox{d}z}{\hbox{d}Z_a\over\hbox{d}z}
+(\epsilon\mu\,\omega^2-k^2)Z_a=0                             \label{car3a}
\end{eqnarray}
with separation constants $k_x$, and $k_y$ meaning physically
wave numbers, and with $k^2=k_x^2+k_y^2$. The solutions of the
first two equations (\ref{car1}-\ref{car2}) are, of course, exponentials
or sines and cosines
\begin{equation}
a=e^{i(k_x x+k_y y)}Z_a\ ,                                  \label{sep}
\end{equation}
but the solutions of (\ref{car3a}) can be weird
and convey physical information never considered before.

According to (\ref{repa1b}), the representative $b$ is subject to
another partial differential equation
\begin{equation}
{\partial^2 b\over \partial x^2}+
{\partial^2 b\over \partial y^2}+
{\partial^2 b\over \partial z^2}-
{\hbox{d}\log{\mu}\over\hbox{d}z}{\partial b\over\partial z}
+\epsilon\mu\omega^2 b=0\ .                                \label{rep2b}
\end{equation}
When this is separated using the ansatz $b=X\,Y\,Z_b$ similar to
(\ref{sepcar}), the ordinary differential equations for $X$ and $Y$
the same as (\ref{car1}) and (\ref{car2}), respectively,
but the differential equation for $Z_b$
\begin{equation}
{\hbox{d}^2 Z_b\over\hbox{d}z^2}
-{\hbox{d}\log{\mu}\over\hbox{d}z}{\hbox{d}Z_b\over\hbox{d}z}
+(\epsilon\mu\,\omega^2-k^2)Z_b=0                              \label{car3b}
\end{equation}
is different from (\ref{car3a}). This reflects a responsiveness of
graded materials to polarization. Examples will be discussed
in the next Sections.

It should never be forgotten that the partial differential
equations (\ref{rep1a}-\ref{rep1b}) hold in any cylindrical
coordinate system. For a less trivial example let us select
elliptic-cylinder coordinates $\xi,\eta,\zeta$ \cite[Sec.1]{Moo61}
\begin{eqnarray}
x=c\,\cosh\xi \cos\eta \\
y=c\,\sinh\xi \sin\eta                                         \label{ell}\\
z=\zeta
\end{eqnarray}
$c$ being a  positive constant.
The ordinary differential equations obtained by separation of
\begin{equation}
a\hbox{ or }b=\Xi\,{\rm H}\,Z_{a\hbox{ or }b}\quad\hbox{with}\quad
    \Xi=\Xi(\xi),\ {\rm H}={\rm H}(\eta),\
    Z_{a\hbox{ or }b}=Z_{a\hbox{ or }b\,\omega}(z)                \label{ell0}
\end{equation}
are
\begin{eqnarray}
{\hbox{d}^2\Xi\over\hbox{d}\xi^2}-(q-k^2 c^2\cosh^2\xi)\Xi=0\ , \label{ell1}\\
{\hbox{d}^2{\rm H}\over\hbox{d}\eta^2}+(q-k^2 c^2\cos^2\eta){\rm H}=0\ .
                                                                \label{ell2}
\end{eqnarray}
The separation constants are here $q$ and $k^2$. The solutions of
these two equations are Mathieu functions \cite{Mei54}.
The third equation was not written because it is identical with (\ref{car3a})
or (\ref{car3b}). The elliptic-cylinder coordinates are especially
interesting as they allow to exactly predict the diffraction by a strip or
a slit, not just by an edge, see \cite[Secs.10,11]{Bro10} and the references
therein. Moreover in the theory presented here, the strip need not to be
homogeneous. So we can devise novel ways to bunching and debunching
of electromagnetic waves.

\subsubsection{Graded reflectors, transmitters and polarizers}
\label{sec:graded}
As a first example of application, consider a medium homogenous
and isotropic for $z<0$:
\begin{equation}
\epsilon=\epsilon_-\ ,\qquad \mu=\mu_-                   \label{low}
\end{equation}
with constant permittivity $\epsilon_-$ and permeability $\mu_-$.
The solutions of the differential equations (\ref{rep2a}) and (\ref{rep2b})
are almost trivial:
\begin{equation}
a_-\hbox{ or }b_-=e^{i(k_x x+k_y y)}
(e^{i\sqrt{\epsilon_-\mu_-\omega^2 -k^2}\,z}+R_{a\hbox{ or }b}\
e^{-i\sqrt{\epsilon_-\mu_-\omega^2 -k^2}\,z})\ .              \label{low1}
\end{equation}
They describe waves of different polarization coming from negative infinity
and being partially reflected at $z=0$. $R_a$ and $R_b$ are complex
constants to fix the strengths and phases of the reflected waves.
These constants have to be determined from the solution of the
boundary-value problem; see below.

For $z>0$ the permittivity is supposed to vary
\begin{equation}
\epsilon={\epsilon_\infty\over 1 - (1-\epsilon_\infty/\epsilon_+)\exp(-z/z_+)}\ ,
 \qquad \mu=\mu_+\ ,                                             \label{upp}
\end{equation}
whereas the permeability is supposed to stay constant $\mu_+$.
The permittivity (\ref{upp}) takes the value $\epsilon_+$ at $z=0$
and increases or decreases with positive slope constant $z_+$ while it
approaches $\epsilon_\infty$ for $z\rightarrow\infty$.

The ordinary differential equations (\ref{car3a}) and (\ref{car3b})
differ considerably. Yet both can be solved in terms
of the hypergeometric function
\begin{eqnarray}
Z_{a\hbox{ or }b}=e^{i\sqrt{\epsilon_\infty\mu_+\omega^2-k^2}\,z}
\phantom{(\beta_{a\hbox{ or }b},\gamma_{a\hbox{ or }b},
(1-\epsilon_\infty/\epsilon_+)\exp(-z/z_+))}                   \nonumber\\
\qquad\qquad\qquad
F(\alpha_{a\hbox{ or }b},\beta_{a\hbox{ or }b},\gamma_{a\hbox{ or }b},
(1-\epsilon_\infty/\epsilon_+)\exp(-z/z_+))\ .                  \label{upp1}
\end{eqnarray}
For $Z_a$ we have to use the parameters
\begin{eqnarray}
\alpha_a=1/2\,(1+\sqrt{1+4k^2 z_+^2})
-i\sqrt{\epsilon_\infty\mu_+\omega^2-k^2}\,z_+\ ,               \nonumber\\
\beta_a=1/2\,(1-\sqrt{1+4k^2 z_+^2})
-i\sqrt{\epsilon_\infty\mu_+\omega^2-k^2}\,z_+\ ,               \label{uppa}\\
\gamma_a=1-2i\sqrt{\epsilon_\infty\mu_+\omega^2-k^2}\,z_+\ ,    \nonumber
\end{eqnarray}
but for $Z_b$
\begin{eqnarray}
\alpha_b=+k z_+ -i\sqrt{\epsilon_\infty\mu_+\omega^2-k^2}\,z_+\ , \nonumber\\
\beta_b=-k z_+ -i\sqrt{\epsilon_\infty\mu_+\omega^2-k^2}\,z_+\ , \label{uppb}\\
\gamma_b=1-2i\sqrt{\epsilon_\infty\mu_+\omega^2-k^2}\,z_+\ .      \nonumber
\end{eqnarray}
The very fact that the parameters in (\ref{uppa}) and (\ref{uppb}) differ
shows that graded materials act discriminatorily towards waves with
different polarization.

There are second solutions of the second-degree equations
(\ref{car3a}) and (\ref{car3b}), but we don't need them here.
Namely
\begin{equation}
F(\alpha,\beta,\gamma,(1-\epsilon_\infty/\epsilon_+)\exp(-z/z_+))
\rightarrow 1\hbox{ for }z\rightarrow\infty                     \label{upp2}
\end{equation}
is a property of the hypergeometric function for all values of
$\alpha,\beta,\gamma$. Therefore (\ref{upp1}) is sufficient
to describe a wave running to positive infinity.

Thus we have, also for $z>0$, solutions of the differential equations
(\ref{rep2a}) and (\ref{rep2b}):
\begin{eqnarray}
a_-\hbox{ or }b_-=e^{i(k_x x+k_y y)}\ T_{a\hbox{ or }b}\
e^{i\sqrt{\epsilon_\infty\mu_+\omega^2-k^2}\,z}
\hbox to 104pt{\hfil}                                            \nonumber\\
F(\alpha_{a\hbox{ or }b},\beta_{a\hbox{ or }b},\gamma_{a\hbox{ or }b},
(1-\epsilon_\infty/\epsilon_+)\exp(-z/z_+))\ .                  \label{upp3}
\end{eqnarray}
$T_a$ and $T_b$ are complex constants to seize the strengths and
phases of the transmitted wave.

The four coefficients $R_a,T_a$ and $R_b,T_b$ are
determined by the conditions at the boundary $S$. The surface $S$
is given here as $z=0$. The normal ${\bf n}$ coincides with
the vector ${\bf e}_z$. Specializing (\ref{boua0}-\ref{boub0}) yields
\begin{eqnarray}
Z_a\,|_{z=-0}=Z_a\,|_{z=+0}\ ,\qquad
{1\over\epsilon}{\hbox{d}Z_a\over\hbox{d}z}\,\bigg|_{z=-0}=
{1\over\epsilon}{\hbox{d}Z_a\over\hbox{d}z}\,\bigg|_{z=+0}\ ,  \label{boua}\\
Z_b\,|_{z=-0}=Z_b\,|_{z=+0}\ ,\qquad
{1\over\mu}{\hbox{d}Z_b\over\hbox{d}z}\,\bigg|_{z=-0}=
{1\over\mu}{\hbox{d}Z_b\over\hbox{d}z}\,\bigg|_{z=+0}\ .       \label{boub}
\end{eqnarray}

The boundary conditions (\ref{boua}) and (\ref{boub}) yield both
two linear equations for $R_a,T_a$ and $R_b,T_b$, respectively, and are
solvable elementarily. Next one differentiates the electromagnetic field
from the representatives (\ref{low1}) and (\ref{upp3})
according to (\ref{sol1}-\ref{sol2}) and calculates
from the electromagnetic field the Pointing vector to obtain
quantities directly comparable to experimental results.

Several modifications and generalizations are at hand. The waves
may run in the opposite directions. In this case the second solutions
of (\ref{car3a}) and (\ref{car3b}) are needed to describe reflected
waves. They too are expressible in terms of the hypergeometric
function. We may also study the transition of waves from one
graded medium to other graded materia. In this case the purely
exponential waves (\ref{low1}) must be replaced with expressions
containing hypergeometric functions.

In the definition of gradation (\ref{upp}), it was assumed that
only the permittivity varies. This is resonable for many materials,
but for independent variations both of permittivity and permeability
it is advisable to use other parametrizations. Power laws, for example,
\begin{equation}
\epsilon=\epsilon_\alpha z^\alpha\ ,\quad
\mu=\mu_\beta z^\beta\quad\hbox{ with }\quad\alpha+\beta=-2,-1,0,2\label{pow}
\end{equation}
lead to solvable differential equations (\ref{car3a}) and (\ref{car3b})
- solvable by functions not more complicated than the confluent
hypergeometric function.

Another useful variability is
\begin{equation}
\epsilon=\epsilon_0\exp{z\over z_\epsilon}\ ,\quad
\mu=\mu_0\exp{z\over z_\mu}                                   \label{tur}
\end{equation}
with positive constants $\epsilon_0,\mu_0$ and positive
or negative constant decay lengths $z_\epsilon$ and $z_\mu$.
Apt solutions of (\ref{car3a}) and (\ref{car3b}) are
Hankel functions $H_\nu^{(1)}$ and $H_\nu^{(2)}$ of weird index
and weird argument:
\begin{eqnarray}
Z_a=\exp\bigg({z\over 2 z_\epsilon}\bigg)\
H_\nu^{(1,2)}\bigg({2 z_\epsilon z_\mu\sqrt{\epsilon_0\mu_0}\,\omega\over
     z_\epsilon+z_\mu}\exp{(z_\epsilon+z_\mu)z\over 2z_\epsilon z_\mu}\bigg)
                                                              \nonumber\\
\hbox{ with }\nu={z_\mu\sqrt{1+4k^2 z_\epsilon^2}\over z_\epsilon+z_\mu}\ ,
                                                              \label{tura}\\
Z_b=\exp\bigg({z\over 2 z_\mu}\bigg)\
H_\nu^{(1,2)}\bigg({2 z_\epsilon z_\mu\sqrt{\epsilon_0\mu_0}\,\omega\over
       z_\epsilon+z_\mu}\exp{(z_\epsilon+z_\mu)z\over 2z_\epsilon z_\mu}\bigg)
                                                              \nonumber\\
\hbox{ with }\nu={z_\epsilon\sqrt{1+4k^2 z_\mu^2}\over z_\epsilon+z_\mu}\ .
                                                              \label{turb}
\end{eqnarray}
Hankel functions, also denoted as Bessel functions of third kind
\cite[Chap.III]{Mag66}, are more appropriate for the description
of running waves than ordinary Bessel and Neumann functions.

The parametrizations (\ref{pow}) or (\ref{tur}) should be applied to slabs
as there is no medium with infinitely large or infinitely small
material properties. Instead of one block of boundary-value conditions
as in (\ref{boua}-\ref{boub}) there should be two blocks, one for a boundary
at, say, $z=z_1$ and the other block for a surface at $z=z_2$. The somewhat
larger linear systems do not essentially aggravate the solution of the entire
problem.

\subsubsection{Stopping the energy flux}
\label{sec:stopping}
A peculiar special case of (\ref{tur}) is
\begin{equation}
\epsilon=\epsilon_0\exp{z\over z_\epsilon}\ ,\quad
\mu=\mu_0\exp{-z\over z_\epsilon}\ .                            \label{tur2}
\end{equation}
The index of refraction (\ref{ind}) is just 1. Believers in the Helmholtz
equation (\ref{hel}) should expect just ordinary propagation of waves,
but no spectacular effect.

The true differential equations
(\ref{car3a}) and (\ref{car3b}) have constant coefficients:
\begin{eqnarray}
{\hbox{d}^2 Z_a\over\hbox{d}z^2}
-{1\over z_\epsilon}{\hbox{d}Z_a\over\hbox{d}z}
+(\epsilon_0\mu_0\omega^2-k^2)Z_a=0\ ,                        \label{tur3a}\\
{\hbox{d}^2 Z_b\over\hbox{d}z^2}
+{1\over z_\epsilon}{\hbox{d}Z_b\over\hbox{d}z}
+(\epsilon_0\mu_0\omega^2-k^2)Z_b=0\ .                        \label{tur3b}
\end{eqnarray}
Hence their solutions are readily found:
\begin{eqnarray}
Z_a=\exp\bigg({z\over 2z_\epsilon}\bigg)
\exp\bigg(\pm\sqrt{{1\over 4z_\epsilon^2}
-(\epsilon_0\mu_0\omega^2-k^2)}\,z\bigg)\ ,
                                                           \label{tur4a}\\
Z_b=\exp\bigg({-z\over 2z_\epsilon}\bigg)
\exp\bigg(\pm\sqrt{{1\over 4z_\epsilon^2}
-(\epsilon_0\mu_0\omega^2-k^2)}\,z\bigg)\ .
                                                           \label{tur4b}
\end{eqnarray}
The exponential functions directly behind the equal signs
impress much, but for the energy flux they don't matter at all.
They only exist to compensate the factors $1/\epsilon$ and $1/\mu$
in (\ref{poi1}).
The astonishing item is the square root in the second exponentials.
Normally, i.e.\ for $z_\epsilon\rightarrow\infty$ as it holds for any almost
homogeneous medium, the value of the root is imaginary and the
exponential function carrying it is complex, just the description
of a plane wave propagating we are used to. Yet if
\begin{equation}
|z_\epsilon|<{1\over 2\sqrt{\epsilon_0\mu_0\omega^2-k^2}}\ ,     \label{tur5}
\end{equation}
the functions (\ref{tur4a}) and (\ref{tur4b}) become real and
the energy flux against the gradation is stopped.
The condition means that the slope constant must be smaller
than the wavelength the wave would have in a medium without gradation
divided by $4\pi$. Though this is a short length, it can be
constructed in modern labs.

Proof: Evaluation of (\ref{poi1}) using (\ref{tur4a}) and (\ref{sep})
yields
\begin{equation}
{\bf S}\propto{\bf e}_x k_x+{\bf e}_y k_y+{\bf e}_z
\cases{\sqrt{\epsilon_0\mu_0\omega^2-k^2-1/(2z_\epsilon)^2}
           &if\ $\epsilon_0\mu_0\omega^2>k^2+1/(2z_\epsilon)^2$ \cr
        0 &otherwise\cr}                                         \label{tur6}
\end{equation}
The result is the same when the electromagnetic wave is represented
by $b$ with $Z_b$ from (\ref{tur4b}).

The most surprising feature of the effect is its independence of the sign
of the slope constant $z_\epsilon$. It doesn't matter if permittivity
increases or decreases. Important is only a sufficiently steep change.

The finding (\ref{tur6}) differs fundamentally from the behavior of
electromagnetic waves in conducting materials. There the energy
intrudes and is dissipated. The finding is also fundamentally
different from the behavior in homogeneous dielectric materials.
Even when there is a discontinuity, part of the wave is maybe reflected,
but the remainder goes on to transport energy.
The waves found here differ as well from the evanescent waves
which make possible dielectric waveguides. In these waveguides,
$k$ must be greater than a positive cut-off wavenumber,
whereas the stopping described here works also for $k=0$.

The effect is moreover not singular. The case (\ref{tur2})
need not be fulfilled exactly. One can derive this from the asymptotic
expansions of the Hankel functions in (\ref{tura}-\ref{turb}). When both
index and argument get great, the Hankel functions become exponential
functions with real argument if the index is greater than the argument,
but they become exponential functions with imaginary argument in the opposite
case \cite[Sec.3.14.2]{Mag66}, similar to the
elementary functions in (\ref{tur4a}-\ref{tur4b}). Generally,
however, there is some dependence of the stopping on polarization.

Time-dependent analysis reveals that the energy density oscillates.
During one half of the period $2\pi/\omega$ it is pushed into the graded
medium, during the other half it is withdrawn. The depth of the penetration
is approximately described by the second exponential functions
in (\ref{tur4a}-\ref{tur4b}) taken with negative signs before the roots.
Therefore, in a slab of finite thickness, the stopping is not perfect.
Waves impinging on the one boundary of the graded medium decrease
in the medium, but the mechanism just described may excite waves,
though weak ones, on the other boundary. Exact amplitudes and phases
follow from the boundary conditions (\ref{boua}-\ref{boub}).

\subsubsection{Dielectric mirrors}
\label{sec:mirror}
Having read the previous Section one may argue that monotonous exponential
growth cannot be maintained on long distances. However, one can realize
similar stopping with zigzagging material properties.
This section is devoted to periodic variations of the permittivity.
Let
\begin{equation}
\epsilon=\epsilon_0-\epsilon_1\cos{k_0 z}\ ,
 \qquad \mu=\mu_0                                                \label{mir}
\end{equation}
with positive constants $\epsilon_0$, $\epsilon_1$, $\mu_0$ and $k_0$.
When we compare the resulting equation (\ref{car3b})
\begin{equation}
{\hbox{d}^2 Z_b\over\hbox{d}z^2}+(\epsilon_0\mu_0\omega^2 - k^2
-\epsilon_1\mu_0\omega^2\cos k_0z )Z_b=0                         \label{mir1}
\end{equation}
to the equation of the Mathieu functions $\me_\nu(w;q)$
\begin{equation}
{\hbox{d}^2 \me_{\pm\nu}(w;q)\over\hbox{d}w^2}+(\lambda
-2q\cos 2w )\me_{\pm\nu}(w;q)=0                           \label{mir2}
\end{equation}
with constant $\lambda$ and $q$
as defined by Meixner and Sch\"afke \cite[p.105]{Mei54},
see \cite[p.404]{Whi73} for a slightly different definition, we find
\begin{equation}
Z_b=\me_{\pm\nu}(k_0 z/2;q),\quad
\lambda=4(\epsilon_0\mu_0\omega^2-k^2)/k_0^2,\quad
q=2\epsilon_1\mu_0\omega^2/k_0^2\ .                             \label{mir3}
\end{equation}

The Mathieu functions possess properties transcending the flexibility
of the hypergeometric function and all its descendants. It follows from
Floquet's theorem \cite[p.412]{Whi73} that one can compute them from
a Fourier series times an exponential factor
\begin{equation}
\me_\nu(w;q)=e^{i\nu w}
         \sum_{n=-\infty}^{n=+\infty}c_{2n}^\nu(q)e^{i2nw}\ .   \label{mir4}
\end{equation}
with coefficients $c_{2n}^\nu(q)$ for which Hill's theory \cite[p.413]{Whi73}
provides handy expressions.

The characteristic exponent $\nu$ is a surprising function of $\lambda$
and $q$. Imagine $q>0$ fixed while $\lambda$ varies. Then $\nu$ assumes,
for certain bands of $\lambda$, only real values.
This is what one should expect for physical reasons: A periodic
perturbation in the differential equation causes periodic or
rather quasi-periodic perturbations in the solutions. However,
in the complements of these bands, $\nu$ acquires complex values. The
phenomenon is known as parametric amplification, but it is often
forgotten that only one solution describes amplification, whereas the other
describes attenuation. This type of damping, which comes about without
friction or spatial dissipation, is the effect we want to consider here.

Floquet's theorem and Hill's theory were reinvented and generalized
in solid-state physics where the theory is known as Bloch's
theorem\cite{Blo28},\cite{Kit96}.
In solid-state physics, $\lambda$ plays the role of the energy in
Schr\"odinger's equation and the complementary bands are denoted
as {\it forbidden\/}.

The lowest forbidden band of Mathieu's equation (\ref{mir2})
is characterized \cite[p.120]{Mei54} by
\begin{equation}
|\lambda-1|<|q|+O(q^2)\ .                                   \label{ban1}
\end{equation}
Maximum attenuation, described by the imaginary part $\Im$ of the
characteristic exponent $\nu$, takes place in its median \cite[p.165]{Mei54}
\begin{equation}
\Im \nu=|q|/2+O(q^2)\quad\hbox{ where }\lambda=1+O(q^2)\ .   \label{ban2}
\end{equation}
To estimate the attenuation of the Pointing vector (\ref{poi1}), we must
take the absolute square of the leading factor on the right-hand side
of (\ref{mir4}). The attenuation of energy flux is thus
\begin{equation}
\bar{\bf S}\propto\exp(-|q|w+O(q^2))\ .                      \label{ban3}
\end{equation}

Applying this to the problem at hand (\ref{mir3}), we see the first
forbidden band approximately defined by
\begin{equation}
|4(\epsilon_0\mu_0\omega^2-k^2)-k_0^2|< 2\epsilon_1\mu_0\omega^2\ .\label{mir5}
\end{equation}
The median of the forbidden band is where the left-hand side is zero.
One may write this as
\begin{equation}
{1\over k_0}\approx {1\over 2\sqrt{\epsilon_0\mu_0\omega^2-k^2}}\label{mir6}
\end{equation}
meaning that the wave length of the intruding light must be twice
the wave length of the dielectric zigzagging. In other words: We must have
four layers of different media or two periods of permittivity for every
spatial period of light. The reader is encouraged to compare (\ref{mir6})
with (\ref{tur5}).

When (\ref{mir6}) is used in the formula for $q$
(\ref{mir3}) and if it is assumed, just for simplicity, that $k=0$, i.e.
the light impinges vertically, we obtain as a crude estimate
\begin{equation}
q\approx {\epsilon_1\over 2\epsilon_0}                          \label{qest}
\end{equation}
and thus for the attenuation (\ref{ban3})
\begin{equation}
\bar{\bf S}\propto\exp\bigg(-{\epsilon_1\over\epsilon_0}{k_0 z\over 4}\bigg)\ .
                                                                  \label{mir7}
\end{equation}
A variation of permittivity $\epsilon_1\approx 0.13\epsilon_0$ appears to be
realistic. Then, according to (\ref{mir7}), it takes less than 6
periods of permittivity, or 11 layers, to attain attenuation by a decade,
it takes less than 23 layers to attain attenuation by two decades and so
forth. An electromagnetic wave that impinges on a periodic
structure cannot penetrate. It is reflected.

The dielectric mirror just described is selective. Light that has not the
suitable wave length (\ref{mir6}) passes. Yet selectiveness
isn't overly sharp. According to (\ref{ban1}), the width of the band
is $2|q|$. If $q$ is estimated according to (\ref{qest}),
we find that the relative width of the forbidden band is
$\epsilon_1/\epsilon_0$, i.e.\ 25\% in the present
example. This is enough to cover a considerable part of the spectrum
visible to human eyes. Moreover the dielectric mirror alters its properties
with the angle of incidence. The angle is contained in the
transverse wave number $k$ and enters the theory via the parameter $\lambda$
in (\ref{mir3}).

It remains to check the dielectric mirror for its dependence on
polarization. To this end the solution of the differential equation
(\ref{car3a}) has to be compared with the solution of (\ref{car3b})
which we just discussed.

The term with the first derivatives in (\ref{car3a}) can be eliminated
introducing the auxiliary function $\tilde Z_a$, yet
at the cost of more complications in the factor of $\tilde Z_a$:
\begin{eqnarray}
\tilde Z_a=Z_a/\sqrt{\epsilon}\ ,
\phantom{{1\over 2\epsilon}{\hbox{d}^2 \epsilon\over\hbox{d}z^2}
-{3\over 4\epsilon^2}\bigg({\hbox{d}\epsilon\over\hbox{d}z}
}\nonumber\\
{\hbox{d}^2 \tilde Z_a\over\hbox{d}z^2}+\bigg(\epsilon\mu\omega^2-k^2
+{1\over 2\epsilon}{\hbox{d}^2 \epsilon\over\hbox{d}z^2}
-{3\over 4\epsilon^2}\bigg({\hbox{d}\epsilon\over\hbox{d}z}\bigg)^2\bigg)
\tilde Z_a=0\ .                                               \label{mir8}
\end{eqnarray}
Generally this is not exactly a Mathieu equation, but it is,
because of its periodic coefficient, of Hill's type.
It can be solved in the same way as Mathieu's
and exhibits the same features, namely allowed and forbidded bands.
Nevertheless for small oscillations of the permittivity
$\epsilon_1\ll\epsilon_0$, equation (\ref{mir8}) can be approximated by the
Mathieu equation
\begin{equation}
{\hbox{d}^2\tilde Z_a\over\hbox{d}z^2}+(\epsilon_0\mu_0\omega^2-k^2
-\bigg(\epsilon_1\mu_0\omega^2-{\epsilon_1\over\epsilon_0}
{k_0^2\over 2}\bigg)\cos k_0z )\tilde Z_a=0\ ,                  \label{mir1a}
\end{equation}
i.e.\ $\tilde Z_a$ is represented by a Mathieu function, too,
where $\lambda$ is same as in (\ref{mir3}), but
\begin{equation}
q={2\epsilon_1\mu_0\omega^2\over k_0^2}-{\epsilon_1\over\epsilon_0}
\approx{-\epsilon_1\over 2\epsilon_0}\ .                        \label{mir3a}
\end{equation}
Repeating the same deliberations as those following (\ref{mir3}), we find
that the median of the forbidden band is at the same position
(\ref{mir5}). The parameter $q$ has now, apart from its sign,
approximately the same magnitude as in (\ref{qest}), but the sign of $q$
can be compensated by an unimportant phase shift of the argument in
Mathieu's function. Therefore both the width of the forbidden band
and the attenuation are approximately the same as for $Z_b$. Thus,
surprisingly enough, the dielectric mirror depends but weakly on polarization.

\subsection{Central-symmetric variations of material properties}
\label{sec:central}

For this case we specialize the condition (\ref{dir}) by
${\bf v}_0={\bf 0}$ and $v_1=1$. The carrier is just the vector
of position ${\bf v}={\bf r}$ and hence ${\bf v}^2=r^2$.
In all equations of Section \ref{sec:reshaping} we got to replace
$\zeta$ with the $r$ customary in the spherical coordinate system
$r,\theta,\varphi$. The differential equations (\ref{rep1a}-\ref{rep1b})
appear as
\begin{eqnarray}
{1\over r}{\partial^2 ra\over\partial r^2}+
{1\over r^2}\bigg({1\over\sin\theta}{\partial\over\partial\theta}
\bigg(\sin\theta{\partial a\over\partial\theta}\bigg)
+{1\over\sin^2\theta}{\partial^2 a\over\partial\varphi^2}\bigg)
-{\hbox{d}\log{\epsilon}\over\hbox{d}r}\,{1\over r}{\partial ra\over\partial r}
+\epsilon\mu\omega^2 a=0,\quad                             \label{rep3a}\\
{1\over r}{\partial^2 rb\over\partial r^2}+
{1\over r^2}\bigg({1\over\sin\theta}{\partial\over\partial\theta}
\bigg(\sin\theta{\partial b\over\partial\theta}\bigg)
+{1\over\sin^2\theta}{\partial^2 b\over\partial\varphi^2}\bigg)
-{\hbox{d}\log{\mu}\over\hbox{d}r}\,{1\over r}{\partial rb\over\partial r}
+\epsilon\mu\omega^2 b=0.\quad                              \label{rep3b}
\end{eqnarray}
Both equations can be separated by similar ansatzes
\begin{eqnarray}
a={1\over r}R_a\,Y_{lm}\quad\hbox{ and }\quad b={1\over r}R_b\,
Y_{lm}\quad\hbox{ with }\quad                                     \nonumber\\
R_a=R_{a\omega}(r),\  R_b=R_{b\omega}(r),\ Y_{lm}=Y_{lm}(\theta,\varphi).
                                                                  \label{cen1}
\end{eqnarray}
The equation for the angular factor $Y_{lm}$ is the same for both
representatives $a$ and $b$:
\begin{equation}
{1\over\sin\theta}{\partial\over\partial\theta}
\bigg(\sin\theta{\partial Y_{lm}\over\partial\theta}\bigg)
+{1\over\sin^2\theta}{\partial^2 Y_{lm}\over\partial\varphi^2}
+l(l+1)Y_{lm}=0\ .                                         \label{cen2}
\end{equation}
It is the differential equation of the familiar spherical harmonics.
They must be unique. So $l$ and $m$ must be integers, in fact $l=1,2,3,...$,
and $|m|\le l$. $l=0$ is excluded because $Y_{00}$ is a constant,
and a representative not depending at all on the angles is annihilated
by the curls in (\ref{sol1}) or (\ref{sol2}).

The differential equations for the radial parts are extraordinary:
\begin{eqnarray}
{\hbox{d}^2 R_a\over\hbox{d}r^2}
-{\hbox{d}\log{\epsilon}\over\hbox{d}r}{\hbox{d}R_a\over\hbox{d}r}
+\bigg(\epsilon\mu\omega^2-{l(l+1)\over r^2}\bigg)R_a=0,      \label{cen3a}\\
{\hbox{d}^2 R_b\over\hbox{d}r^2}
-{\hbox{d}\log{\mu}\over\hbox{d}r}{\hbox{d}R_b\over\hbox{d}r}
+\bigg(\epsilon\mu\omega^2-{l(l+1)\over r^2}\bigg)R_b=0.      \label{cen3b}
\end{eqnarray}
Although they look like their one-dimensional analogs in (\ref{car3a})
and (\ref{car3b}), the last terms on the left-hand sides are
different. Nevertheless also the present differential equations
can be solved for all power laws conforming to
\begin{equation}
\epsilon=\epsilon_\alpha r^\alpha\ ,\quad
\mu=\mu_\beta r^\beta\quad\hbox{ with }\quad\alpha+\beta=-2,-1,0,2\label{powc}
\end{equation}
i.e.\ solved in terms of functions not more complicated than the confluent
hypergeometric function.

\subsubsection{Bound electromagnetic waves}
\label{sec:bound}
In Schr\"odinger's quantum mechanics, electrons can be bound
in a spherical well. The eigenvalues of Schr\"odinger's
solutions are discrete. A similar construction for photons isn't known.
For instance
\begin{equation}
\epsilon=\cases{\epsilon_0  &if\ $0\le r< r_1$,\cr
                \epsilon_1  &if\ $r_1\le r <\infty$.\cr}\qquad
\mu=\cases{\mu_0  &if\ $0\le r< r_1$,\cr
           \mu_1  &if\ $r_1\le r <\infty$.\cr}                    \label{bou1}
\end{equation}
with positive constants $\epsilon_0,\mu_0$ inside the spherical core
$r<r_1$ and other positive constants $\epsilon_1,\mu_1$ outside
admits as solution of (\ref{cen3a}) only
\begin{equation}
{1\over r}R_a=
\cases{A\,j_l(\sqrt{\epsilon_0\mu_0}\,\omega r)
                 &if\ $0\le r< r_1$,\cr
B\,h_l^{(1)}(\sqrt{\epsilon_1\mu_1}\,\omega r)+
C\,h_l^{(2)}(\sqrt{\epsilon_1\mu_1}\,\omega r)
                 &if\ $r_1\le r <\infty$.\cr}                     \label{bou2}
\end{equation}
with spherical Bessel functions $j_l$, spherical Hankel functions
$h_l^{(1)},h_l^{(1)}$, $l=1,2,3\ldots$ \cite[Sec.10]{Abr70}, and
constants $A$, $B$ and $C$. All functions describe running waves.
One might think of materials where the real permittivity or
the permeability are negative. Such materials exist,
but those negative values take place only in narrow bands of $\omega$
and come always with considerable conductivity. So there is not the
least chance to establish bound electromagnetic waves and discrete
values of $\omega$ with spatially constant material properties.

With graded materials, however, we can construct a home of bound waves.
Consider instead of (\ref{bou1})
\begin{equation}
\epsilon=\cases{\epsilon_0       &if\ $0\le r< r_1$,\cr
                \epsilon_1 r_1/r &if\ $r_1\le r <\infty$.\cr}\qquad
\mu=\cases{\mu_0       &if\ $0\le r< r_1$, \cr
           \mu_1 r_1/r &if\ $r_1\le r <\infty$. \cr}              \label{bou3}
\end{equation}
The respective solution of (\ref{cen3a}) is
\begin{equation}
R_a=\cases{A\,r\,j_l(\sqrt{\epsilon_0\mu_0}\,\omega r)
                 &if\ $0\le r< r_1$,\cr
           B\,r^{-\sqrt{l(l+1)-\epsilon_1\mu_1\omega^2 r_1^2}}
                 &if\ $r_1\le r <\infty$.\cr}                     \label{bou4}
\end{equation}
The solution on the flank of the wall $r_1\le r$ can be a wave,
though a weird one
\begin{equation}
r^{\pm\sqrt{l(l+1)-\epsilon_1\mu_1\omega^2 r_1^2}}=
\exp({\pm i \sqrt{\epsilon_1\mu_1\omega^2 r_1^2-l(l+1)}\log r}) \label{bou5}
\end{equation}
if $\omega$ is sufficiently high. Yet if
\begin{equation}
\Omega=\sqrt{\epsilon_0\mu_0}\,\omega r_1<
\sqrt{{\epsilon_0\mu_0\over\epsilon_1\mu_1}l(l+1)}              \label{bou6}
\end{equation}
with $\Omega$ as nondimensional substitute of $\omega$,
the flank function in (\ref{bou4}) just decreases without any
variation of phase such that the same considerations apply as in Section
\ref{sec:stopping}: The transfer of energy through the flanks is stopped.
One can check this explicitly evaluating the Pointing vector (\ref{poi1})
with (\ref{bou4}) and (\ref{cen1}).

To find the eigenvalues of $\Omega$ and thus of $\omega$,
we must satisfy the boundary conditions (\ref{boua0}).
The surface $S$ is now the sphere $r=r_1$ and the
normal vector is ${\bf n}={\bf r}/r$. Hence
\begin{eqnarray}
R_a\,|_{r=r_1-0}=R_a\,|_{r=r_1+0}\ ,\qquad
{1\over\epsilon}{\hbox{d}R_a\over\hbox{d}r}\,\bigg|_{r=r_1-0}=
{1\over\epsilon}{\hbox{d}R_a\over\hbox{d}r}\,\bigg|_{r=r_1+0}\ .\label{bou7}
\end{eqnarray}
This produces a homogenous linear system for $A$ and $B$. It has a
non-trivial solution if its determinant is zero:
\begin{eqnarray}
\sqrt{\epsilon_0\mu_1\over \epsilon_1\mu_0}
\sqrt{{\epsilon_0\mu_0\over \epsilon_1\mu_1}l(l+1)-\Omega^2}
=-{(\Omega j_l(\Omega))'\over j_l(\Omega)}                       \label{bou8}
\end{eqnarray}
the prime indicating differentiation with respect to the argument $\Omega$.
The function on the left-hand side is a parabola open to the left.
We need its positive branch at positive values of $\Omega$. The parabola
disappears for $\Omega$ greater than
the cut-off given on the right-hand side of (\ref{bou6}).
The function on the ride-hand side of (\ref{bou8})
takes the value $-(l+1)$ at $\Omega=0$. It increases with $\Omega$
and crosses the $\Omega$-axis at the zero of $(\Omega j_l(\Omega))'$.
The function continues to increase until it approaches its pole at the zero
of $j_l(\Omega)$. To secure the existence of a solution of (\ref{bou8}),
it would be sufficient to demand that the cut-off in (\ref{bou6}) be greater
than the zero of $(\Omega j_l(\Omega))'$. But these zeros are not tabulated.
So let us be generous and demand that the cut-off be greater
than the first zero of $j_l(\Omega)$. This yields a condition
\begin{equation}
\sqrt{\epsilon_0\mu_0\over\epsilon_1\mu_1}>{\Omega_l\over\sqrt{l(l+1)}}
\quad\hbox{ where }j_l(\Omega_l)=0\                             \label{rati}
\end{equation}
which warrants the existence of at least one positive solution of
(\ref{bou8}). The expression on the right-hand side tends to 1 as
$l$ tends to infinity. Therefore the restriction on the ratio of
the indices of refraction is unimportant at high multipolarities.
Yet even for $l=1$ the condition (\ref{rati}) can be fulfilled.
The first zero of the first spherical Bessel function is
$\Omega_1\approx 4.5$ \cite[Sec.10]{Abr70} such that ratios of the
indices must be greater than 3.2. In modern times where indices
of refraction can be made as big as 38.6 \cite{Cho11}, this is
moderate requirement.

Waves of the other polarization can found by replacing the
representative $a$ with $b$ and by an interchange of $\epsilon$
and $\mu$. The characteristic equation of this case differs from
(\ref{bou8}) just by a different leading factor. Therefore
it depends on the polarization whether an electromagnetic wave
can be bound, but it does not depend much.

One might compare the construction explained here with a hydrogen atom.
Rather it is similar to a nucleon bound in a collective nuclear
potential as the spectrum of eigenvalues is finite. The essential
difference, however, is that every electron always carries the same
charge which necessitates a normalization of its wave.
Here, by contrast, the energy of the bound electromagnetic
wave is arbitrary. The only necessity to confine the energy is a possible
breakdown of material properties (\ref{resp1}-\ref{resp3}).
It is therefore blameworthy to speak about {\lq\lq}photonics{\rq\rq},
an {\lq\lq}atom for photons{\rq\rq} and so forth. Nevertheless,
if quantum electrodynamics were true, the energy stored in the
construction just described should be discrete.

As in Section \ref{sec:stopping} people might argue that the system
just constructed is not realistic. Fortunately there is no
singularity at the origin at $r=0$, but it is certainly questionable
to require permittivity and permeability approaching zero
as in (\ref{bou3}). The solution of this problem, however, is known.
One must replace the monotonous decrease with zigzagging as explained
in Section \ref{sec:mirror}. This will work. For the differential equations
(\ref{car3a}) and (\ref{cen3a}) are the same for $r\rightarrow\infty$.

\section{An alternative theorem of representation}
\label{sec:alternative}
In the study of graded fibers, one cannot use the
theorem of representation provided in Section \ref{sec:triple}.
It is possible to analyze electromagnetic fields in cylindrical bodies,
but the material properties must not vary except in the direction of the axis
of the cylinder. For graded waveguides, one needs permittivity and permeability
varying with the distance from the axis, i.e.\
$\epsilon=\epsilon_\omega(\rho)$, $\mu=\mu_\omega(\rho)$
in circular cylindrical coordinates $\rho,\varphi,z$.
Yet there is no carrier according to (\ref{dir})
that would be proportional to $\nabla\epsilon$ and
$\nabla\mu$. Fortunately we can rely on the

\medskip
\noindent{\bf Two-Dimensional Representation Theorem.}
{\sl In a system of orthogonal coordinates $\xi,\eta,\zeta$ where
the elements of the metric tensor
\begin{equation}
g_{\xi\xi}=g_{\xi\xi}(\eta,\zeta),\quad
g_{\eta\eta}=g_{\eta\eta}(\eta,\zeta),\quad
g_{\zeta\zeta}=g_{\zeta\zeta}(\eta,\zeta),                 \label{met}
\end{equation}
cf.\ the line element (\ref{line}), do not depend on the
{\it distinguished\/} coordinate $\xi$ and where permittivity
and permeability
\begin{equation}
\epsilon=\epsilon_\omega(\eta,\zeta),\quad
\mu=\mu_\omega(\eta,\zeta)                                 \label{mat}
\end{equation}
do not depend on $\xi$, the fields
\begin{eqnarray}
{\bf E}={1\over\epsilon}\nabla\times{\bf e}_\xi{a\over\sqrt{g_{\xi\xi}}}
        -i\omega{\bf e}_\xi{b\over\sqrt{g_{\xi\xi}}}             \label{solra}\\
{\bf H}=-i\omega{\bf e}_\xi{a\over\sqrt{g_{\xi\xi}}}
       -{1\over\mu}\nabla\times{\bf e}_\xi{b\over\sqrt{g_{\xi\xi}}}\label{solrb}
\end{eqnarray}
solve Maxwell's equations (\ref{max1a}-\ref{max4a}) including
the constitutive relations (\ref{con1}-\ref{con2}) if the
representatives $a$ and $b$
\begin{equation}
a=a_\omega(\eta,\zeta),\quad b=b_\omega(\eta,\zeta)           \label{repr}
\end{equation}
do not depend on $\xi$ and obey the differential equations
\begin{eqnarray}
\sqrt{g_{\xi\xi}\over g_{\eta\eta}g_{\zeta\zeta}}
\bigg[
\bigg({\partial\over\partial\eta}
\sqrt{g_{\zeta\zeta}\over g_{\eta\eta}g_{\xi\xi}}
{\partial a\over\partial\eta}\bigg) +
\bigg({\partial\over\partial\zeta}
\sqrt{g_{\eta\eta}\over g_{\zeta\zeta}g_{\xi\xi}}
{\partial a\over\partial\zeta}\bigg)
\bigg]      \nonumber \\
-{1\over g_{\eta\eta}}{\partial\log{\epsilon}\over\partial\eta}
{\partial a\over\partial\eta}
-{1\over g_{\zeta\zeta}}{\partial\log{\epsilon}\over\partial\zeta}
{\partial a\over\partial\zeta}
+ \epsilon\mu\,\omega^2 a = 0\ ,                               \label{repra}\\
\sqrt{g_{\xi\xi}\over g_{\eta\eta}g_{\zeta\zeta}}
\bigg[
\bigg({\partial\over\partial\eta}
\sqrt{g_{\zeta\zeta}\over g_{\eta\eta}g_{\xi\xi}}
{\partial b\over\partial\eta}\bigg) +
\bigg({\partial\over\partial\zeta}
\sqrt{g_{\eta\eta}\over g_{\zeta\zeta}g_{\xi\xi}}
{\partial b\over\partial\zeta}\bigg)
\bigg]      \nonumber \\
-{1\over g_{\eta\eta}}{\partial\log{\mu}\over\partial\eta}
{\partial b\over\partial\eta}
-{1\over g_{\zeta\zeta}}{\partial\log{\mu}\over\partial\zeta}
{\partial b\over\partial\zeta}
+ \epsilon\mu\,\omega^2 b = 0\ .                             \label{reprb}
\end{eqnarray}}

For the proof, let us start with the representative $a$ only. The ansatz
\begin{eqnarray}
{\bf C}=\nabla\times{\bf e}_\xi{a\over\sqrt{g_{\xi\xi}}}     \label{ansa1}\\
{\bf H}=-i\omega{\bf e}_\xi{a\over\sqrt{g_{\xi\xi}}}         \label{ansa2}
\end{eqnarray}
is the special case of (\ref{solra}-\ref{solrb}) with $b=0$. It
solves at once the Maxwell equations (\ref{max3a}-\ref{max4a}).
Yet under the geometrical restrictions (\ref{met}), (\ref{mat}) and
(\ref{repr}), it also solves Maxwell's equation (\ref{max2a}), namely
\begin{equation}
\nabla{\bf B}={-i\omega\over\sqrt{g_{\xi\xi}g_{\eta\eta}g_{\zeta\zeta}}}
{\partial\sqrt{g_{\eta\eta}g_{\zeta\zeta}}\,\mu\,a\over\partial\xi}=0\label{div}
\end{equation}
because there is nothing behind the differentiation depending on $\xi$.

Thus the only Maxwell equation that still expects solution is (\ref{max1a}).
Using the ansatz (\ref{ansa1}-\ref{ansa2}) it is transformed to
\begin{equation}
\epsilon\nabla\times{1\over\epsilon}\nabla\times{\bf e}_\xi
{a\over\sqrt{g_{\xi\xi}}}
=\epsilon\mu\omega^2\,{\bf e}_\xi{a\over\sqrt{g_{\xi\xi}}}      \label{proof}
\end{equation}
which is equivalent, as we will see soon, to the differential equation
(\ref{repra}). It is apparent that the right-hand side of (\ref{proof})
is proportional to ${\bf e}_\xi/\sqrt{g_{\xi\xi}}$. We will
discover that the same is true for the left-hand side.
To this end we sever the differentiation of $\epsilon$ in (\ref{proof}):
\begin{equation}
\epsilon\nabla\times{1\over\epsilon}\nabla\times{\bf e}_\xi
{a\over\sqrt{g_{\xi\xi}}}
=-(\nabla\log\epsilon)\times\nabla\times{\bf e}_\xi{a\over\sqrt{g_{\xi\xi}}}
+\nabla\times\nabla\times{\bf e}_\xi{a\over\sqrt{g_{\xi\xi}}}\ .
                                                         \quad\label{proof1}
\end{equation}
The nabla operator applied to $\log\epsilon$ yields
\begin{equation}
-\nabla\log\epsilon=
-{{\bf e}_\eta\over\sqrt{g_{\eta\eta}}}
{\partial\log\epsilon\over\partial\eta}
-{{\bf e}_\zeta\over\sqrt{g_{\zeta\zeta}}}
{\partial\log\epsilon\over\partial\zeta}\ .                  \label{proof2}
\end{equation}
The simple curl in (\ref{proof1}) is
\begin{equation}
\nabla\times{\bf e}_\xi{a\over\sqrt{g_{\xi\xi}}}=
{{\bf e}_\eta\over\sqrt{g_{\zeta\zeta}g_{\xi\xi}}}
{\partial a\over\partial\zeta}
-{{\bf e}_\zeta\over\sqrt{g_{\eta\eta}g_{\xi\xi}}}
{\partial a\over\partial\eta}\ .                             \label{proof3}
\end{equation}
Evaluating the cross product of (\ref{proof2}) and (\ref{proof3}) as
required in (\ref{proof1}) produces the second contribution to
(\ref{repra}) times ${\bf e}_\xi/\sqrt{g_{\xi\xi}}$.
The double curl on the right-hand side of (\ref{proof1}) gives the
first contribution to (\ref{repra}) times ${\bf e}_\xi/\sqrt{g_{\xi\xi}}$.
So it is shown that solution of (\ref{repra}) completes the solution
of Maxwell's equations.

The truth of (\ref{reprb}) can be proven when we start from the ansatz
(\ref{solra}-\ref{solrb}) with the representative $b$ only, putting $a=0$.
All Maxwell equations turn out to be automatically solved except (\ref{max3a}).
This one is evaluated as described in (\ref{proof}-\ref{proof3})
where $a$ is interchanged with $b$ and $\epsilon$ with $\mu$.

Finally we remember the linearity of Maxwell's equations. The full
proof of the alternative theorem of representation is just the superposition
of the two proofs produced in the last paragraphs. Q.E.D.

The consistent setup of boundary-value problems is described in the

\noindent{\bf Corollary on boundary-value conditions.} {\sl Let $S$ denote
the line where different media meet, {\bf n} the normal on this line
with ${\bf n}{\bf e}_\xi=0$ and $\partial/\partial n$ the differentiation
along this normal. The representatives $a$ and $b$ must satisfy
\begin{eqnarray}
a\,|_{S-}=a\,|_{S+}\ ,\qquad
{1\over\epsilon}{\partial a\over\partial n}\,\bigg|_{S-}=
{1\over\epsilon}{\partial a\over\partial n}\,\bigg|_{S+}\ ,\label{boua2}\\
b\,|_{S-}=b\,|_{S+}\ ,\qquad
{1\over\mu}{\partial b\over\partial n}\,\bigg|_{S-}=
{1\over\mu}{\partial b\over\partial n}\,\bigg|_{S+}\ .     \label{boub2}
\end{eqnarray}
The symbols $S-$ and $S+$ indicate that the values of the functions
and their derivatives are to be calculated via an approach on the
one side of $S$, say, the low side $S-$, or on the other side, say, the high
side $S+$.}

The proof is nearly the same as the proof of the corollary on boundary-value
conditions in Section \ref{sec:triple}. The main difference is: We must
use now the representation formulae (\ref{solra}-\ref{solrb}).

The reader is kindly asked not to misunderstand the denotation
{\lq\lq}two-dimensional{\rq\rq}. A propagating electromagnetic field
always spans the three-dimensional space. {\lq\lq}Two-dimensional{\rq\rq}
means just that all components of the field depend only on two coordinates.

Yet quite a few problems can be declared to be two-dimensional
by a judicious choice of coordinates. In all these cases it is
advantageous to apply the two-dimensional representation theorem.
Namely the calculation of the electromagnetic field using
(\ref{solra}) and (\ref{solrb}) takes less work than using
(\ref{sol1}) and (\ref{sol2}) as two curls less need to be computed.

When permittivity and permeability are constant in space, the two-dimensional
theorem of representation doesn't offer anything which is not included
in the three-dimensional theorem given in Section \ref{sec:triple}. With
constant material properties, the theorem already presented in
\cite[Sec.10]{Bro10} grants the best systematic approach.

\subsection{Examples in a plane}
\label{sec:plane}

The most straightforward applications of the foregoing theorem take
place in cartesian coordinates $x,y,z$. None of the components of the
metric tensor depends on any coordinate:
\begin{equation}
g_{xx}=1,\quad g_{yy}=1,\quad g_{zz}=1\ .                      \label{metc}
\end{equation}
As distinguished coordinate $\xi$ we may select either $x$ or $y$
or $z$. Let us identify $\xi=z,\eta=x,\zeta=y$. The partial differential
equations (\ref{repra}-\ref{reprb}) appear as
\begin{eqnarray}
{\partial^2 a\over\partial x^2}+{\partial^2 a\over\partial y^2}
-{\partial\log\epsilon\over\partial x}{\partial a\over\partial x}
-{\partial\log\epsilon\over\partial y}{\partial a\over\partial y}
+\epsilon\mu\,\omega^2 a=0\ ,                             \label{reprac}\\
{\partial^2 b\over\partial x^2}+{\partial^2 b\over\partial y^2}
-{\partial\log\mu\over\partial x}{\partial b\over\partial x}
-{\partial\log\mu\over\partial y}{\partial b\over\partial y}
+\epsilon\mu\,\omega^2 b=0\ .                              \label{reprbc}
\end{eqnarray}
The electromagnetic field is represented according to
(\ref{solra}-\ref{solrb}) through
\begin{eqnarray}
{\bf E}={1\over\epsilon}\nabla\times{\bf e}_z a
        -i\omega{\bf e}_z b\phantom{-}                     \label{solrac}\\
{\bf H}=-i\omega{\bf e}_z a -
         {1\over\mu}\nabla\times{\bf e}_z b                \label{solrbc}
\end{eqnarray}
showing that the electromagnetic wave extends in three dimensions
while the representatives $a=a_\omega(x,y)$ and $b=b_\omega(x,y)$
depend on two coordinates $x$ and $y$. Moreover
it should be noticed that both permittivity and permeability may depend
on both coordinates $x$ and $y$. Therefore for the so-called
two-dimensional problem, the equations (\ref{reprac}-\ref{reprbc}) constitute
the most general reduction of the coupled Maxwellian system to two
uncoupled equations.

Even when the partial differential equations (\ref{reprac}-\ref{reprbc})
are not separable, they vastly simplify the solution of Maxwell's
equations as all methods
which people learn in the ordinary courses of quantum mechanics can
be applied directly, for example, Born's approximation and the JWKB,
denoted also as semi-classical approximation. In the latter case, however,
the classical eikonal equation (\ref{eik}) will turn out to be only
of restricted usefulness. Also numerical methods will profit
from the reduction.

When $\epsilon$ and $\mu$ depend only on one coordinate, $x$ or
$y$, the partial differential equations (\ref{reprac}-\ref{reprbc})
can be separated and produce ordinary differential equations
similar to (\ref{car3a}) and (\ref{car3b}). Also when $\epsilon$ and $\mu$
are products of functions which depend on one coordinate only, separation
is possible, but only under certain circumstances. We will see an
example below.

In cartesian coordinates we can choose $x$, $y$ or $z$ as the distinguished
coordinate $\xi$, but a changed choice does not alter the
geometrical situation. In circular cylindrical coordinates $\rho,\varphi,z$
the elements of the metric tensor depend neither on $z$ nor on $\varphi$:
\begin{equation}
g_{\rho\rho}=1,\quad g_{\varphi\varphi}=\rho^2,\quad g_{zz}=1.  \label{metz}
\end{equation}
Hence we may select either $\varphi$ or $z$ as distinguished coordinate, but
now the choice varies the geometrical situation.

Begin with $\xi=z$ as distinguished coordinate.
The partial differential equation (\ref{repra}) appears as
\begin{equation}
{1\over\rho}
\bigg({\partial\over\partial\rho}\rho{\partial a\over\partial\rho}\bigg)+
{1\over\rho^2}{\partial^2 a\over\partial\varphi^2}
-{\partial\log\epsilon\over\partial\rho}{\partial a\over\partial\rho}
-{1\over\rho^2}
{\partial\log\epsilon\over\partial\varphi}{\partial a\over\partial\varphi}
+\epsilon\mu\,\omega^2 a=0                                \label{reprac1}
\end{equation}
The equation (\ref{reprb}) for $b$ is up to an interchange of $\epsilon$
with $\mu$ identical and is therefore not written.

The partial differential equation (\ref{reprac1}) is the equation
of plane scattering. It can be separated
if $\epsilon$ and $\mu$ are functions of $\rho$ only. The ansatz
\begin{equation}
a={\rm P}_a\,\Phi\quad\hbox{with}\quad{\rm P}_a={\rm P}_{a\omega}(\rho)\ ,
\Phi=\Phi(\varphi)                                             \label{cyl0}
\end{equation}
generates the ordinary differential equations
\begin{eqnarray}
{\hbox{d}^2 \Phi\over\hbox{d}\varphi^2}+m^2\Phi=0              \label{cyl1}\\
{\hbox{d}^2{\rm P}_a\over\hbox{d}\rho^2}
+\bigg({1\over\rho}-{\hbox{d}\log{\epsilon}\over\hbox{d}\rho}\bigg)
 {\hbox{d}{\rm P}_a\over\hbox{d}\rho}
+\bigg(\epsilon\mu\,\omega^2-{m^2\over\rho^2}\bigg){\rm P}_a=0  \label{cyl2}
\end{eqnarray}
The first equation has the familiar solutions $\Phi=\exp(i m\varphi)$.
The separation constant $m^2$ must be the square of an integer
$m=0,\pm 1,\pm 2,\pm 3,\ldots$. Otherwise $a$ and hence the
electromagnetic field would not be unique. The second equation can be
solved using functions not more complicated than the confluent
hypergeometric function if
\begin{equation}
\epsilon=\epsilon_\alpha \rho^\alpha\ ,\quad
\mu=\mu_\beta \rho^\beta\quad
                      \hbox{ with }\quad\alpha+\beta=-2,-1,0,2  \label{pows}
\end{equation}
and with constant $\alpha$, $\beta$, $\epsilon_\alpha$ and $\mu_\beta$.
This together with the boundary conditions (\ref{boua2}) gives ample
freedom to model graded centers of scattering around $\rho=0$.

Interestingly the partial differential equation (\ref{reprac1})
can also be separated if permittivity and permeability depend on both
variables, e.g.
\begin{equation}
\epsilon=\rho^\alpha\epsilon_\varphi,\ \mu=\rho^\beta\mu_\varphi
\hbox{ with }\alpha+\beta=-2\hbox{ and }
\epsilon_\varphi=\epsilon_{\varphi\omega}(\varphi),\
\mu_\varphi=\mu_{\varphi\omega}(\varphi).                       \label{doub}
\end{equation}
The functions $\epsilon$ and $\mu$ must be periodic for the uniqueness,
but else they are arbitrary. The ansatz
\begin{equation}
a={\rm P}_a\,\Phi_a\quad\hbox{with}\quad{\rm P}_a={\rm P}_a(\rho),\
\Phi_a=\Phi_{a\omega}(\varphi)                                    \label{doub1}
\end{equation}
produces
\begin{eqnarray}
{\hbox{d}^2\Phi_a\over\hbox{d}\varphi^2}
-{\hbox{d}\log\epsilon_\varphi\over\hbox{d}\varphi}
{\hbox{d}\Phi_a\over\hbox{d}\varphi}
+(q+\epsilon_\varphi\mu_\varphi\omega^2)\Phi_a=0\ ,              \label{cyl3}\\
{\hbox{d}^2{\rm P}_a\over\hbox{d}\rho^2}
+{1-\alpha\over\rho}{\hbox{d}{\rm P}_a\over\hbox{d}\rho}
-{q\over\rho^2}\,{\rm P}_a=0\ .                                  \label{cyl4}
\end{eqnarray}
with the separation constant $q$.
The latter equation is now of Eulerian type simply solved by
\begin{equation}
{\rm P}_a=\rho^{\alpha/2\pm\sqrt{q+\alpha^2/4}}\ ,               \label{cyl5}
\end{equation}
whereas the former equation is of Hill's type. When the average values
of $\epsilon_\varphi$ and $\mu_\varphi$ are denoted by
$\bar{\epsilon}_\varphi$ and $\bar{\mu}_\varphi$, respectively, and if
$\epsilon_\varphi$ and $\mu_\varphi$ oscillate around their average values
but weakly, then (\ref{cyl3}) is solved by periodic Mathieu functions and the
separation constant can be estimated as
\begin{equation}
q\approx m^2-\bar{\epsilon}_\varphi\bar{\mu}_\varphi\omega^2     \label{cyl6}
\end{equation}
$m=0,\pm 1,\pm 2,\pm 3,\ldots$
Because of (\ref{cyl5}) this determines whether we see a wave or
monotonous variation along $\rho$.

There is no sizeable difficulty to solve (\ref{cyl4}) even if the
oscillations of $\epsilon_\varphi$ and $\mu_\varphi$ are large. Use,
for example, Hill's theory.

Presently the devices to simulate invisibility cloaks are mostly plain.
Therefore this is the Section with the best formulas to design them.
The best formulas to design central-symmetrical cloaks can be found in
Section \ref{sec:central}.

\subsection{Examples around an axis}
\label{sec:axis}
When we stay with circular cylindrical coordinates, but select $\xi=\varphi$
as distinguished coordinate, we obtain from (\ref{repra}-\ref{reprb})
\begin{eqnarray}
\rho
\bigg({\partial\over\partial\rho}{1\over\rho}{\partial a\over\partial\rho}\bigg)
+{\partial^2 a\over\partial z^2}
-{\partial\log\epsilon\over\partial\rho}{\partial a\over\partial\rho}
-{\partial\log\epsilon\over\partial z}{\partial a\over\partial z}
+\epsilon\mu\,\omega^2 a=0\ ,                                \label{repraa}\\
\rho
\bigg({\partial\over\partial\rho}{1\over\rho}{\partial b\over\partial\rho}\bigg)
+{\partial^2 b\over\partial z^2}
-{\partial\log\mu\over\partial\rho}{\partial b\over\partial\rho}
-{\partial\log\mu\over\partial z}{\partial b\over\partial z}
+\epsilon\mu\,\omega^2 b=0\ .                                \label{reprbb}
\end{eqnarray}
The reader might notice that the second order operator in these equations
cannot be understood as a part of the Laplace operator. Also one needs
to get used to the equations of representation
\begin{eqnarray}
{\bf E}={1\over\epsilon}
        \bigg(-{{\bf e}_\rho\over\rho}{\partial a\over\partial z}
        +{{\bf e}_z\over\rho}{\partial a\over\partial\rho}\bigg)
        -i\omega{{\bf e}_\varphi\over\rho}\,b                 \label{solraa}\\
{\bf H}=-i\omega{{\bf e}_\varphi\over\rho}\,a
+{1\over\mu}
\bigg({{\bf e}_\rho\over\rho}{\partial b\over\partial z}
      -{{\bf e}_z\over\rho}{\partial b\over\partial\rho}\bigg)\label{solrbb}
\end{eqnarray}
which follow, despite of their weird appearance, directly from (\ref{solra}-\ref{solrb}). For $b=0$
the magnetic field forms rings around the axis $z=0$, for $a=0$ it is the
electric field which clings to circular lines.

Most optical instruments are centered around an axis. The
equations (\ref{repraa}-\ref{solrbb}) provide a better foundation
to design them than anything known up to now. First, their solutions
yield exact solutions of Maxwell's equations. Second, one has to solve
partial differential equations only for one unknown. This is much simpler
than solving the multiply coupled Maxwell equations.
One may, for example, insert a singularity
$a\hbox{ or }b\sim\log((\rho-\rho_o)^2+(z-z_o)^2$
at a point of an object $\rho_o$, $z_o$ and compute where other
singularities arise. One may
simulate lenses by regions in the $\rho-z$ plane where permittivity
$\epsilon$ and permeability $\mu$ are increased. At the same time,
one may similate metallic stops of finite thickness by regions
with complex $\epsilon$, cf. (\ref{crespi}), and study the interaction
between lenses and stops. Many properties of imaging can thus be predicted,
however, with an important exception: astigmatism can not be observed
because the dependence on the azimuth $\varphi$ is missing.

The first example is the graded mono-mode fiber. There we
have permittivity $\epsilon=\epsilon_\omega(\rho)$ and
permeability $\mu=\mu_\omega(\rho)$ as functions of the axial distance
$\rho$ only. Both partial differential equations (\ref{repraa}) and
(\ref{reprbb}) can be separated. Let us select the first for example.
The ansatz with leading $\rho$
\begin{equation}
a=\rho\,{\rm P}_a\,Z\quad\hbox{with}\quad
{\rm P}_a={\rm P}_{a\omega}(\rho),\ Z=Z(z)                \label{cyl7a}
\end{equation}
is advantageous because it
permits simple boundary conditions for $\rho\rightarrow 0$ and
$\rho\rightarrow\infty$, namely ${\rm P}_a=0$ in both cases. Why? We just
have to consider the electromagnetic field in (\ref{solraa}-\ref{solrbb})
and to demand that all its components should stay finite
on the axis and decrease towards infinity. The ansatz (\ref{cyl7a})
generates these ordinary differential equations:
\begin{eqnarray}
{\hbox{d}^2 Z\over\hbox{d}z^2}+k^2 Z=0                        \label{cyl7}\\
{\hbox{d}^2{\rm P}_a\over\hbox{d}\rho^2}
+\bigg({1\over\rho}-{\hbox{d}\log\epsilon\over\hbox{d}\rho}\bigg)
{\hbox{d}{\rm P}_a\over\hbox{d}\rho}
+\bigg(\epsilon\mu\,\omega^2-k^2-{1\over\rho^2}
-{1\over\rho}{\hbox{d}\log\epsilon\over\hbox{d}\rho}\bigg){\rm P}_a=0
                                                                 \label{cyl8}
\end{eqnarray}
The relation between the real wavenumber $k$ and the frequency $\omega$,
the so-called dispersion relation, is the desired item. It is found from
the solution of (\ref{cyl8}) satisfying the boundary conditions
for $\rho\rightarrow 0$ and for $\rho\rightarrow\infty$ as explained
in the previous paragraph. One can solve the ordinary differential equation
(\ref{cyl8}) for permittivities and permeabilities obeying power laws
as in (\ref{pows}) or one can solve it numerically. A boundary-value
problem with one ordinary differential equation is by orders of magnitude
simpler than the same problem with the Maxwell equations and it is much
simpler to attain high accuracy.

The considerations produced here for circular cylindrical coordinates
can be transferred to all coordinate system which embody an axis of
rotation, e.g.\ spherical coordinates, prolate and oblate spheroidal
coordinates, parabolic coordinates and all rotational systems
\cite[secs.I,IV]{Moo61}. Especially interesting appear at first glance
the oblate spheriodal coordinates because they allow an easy study
what an electromagnetic wave does in a bottleneck, and the toroidal
coordinates because they allow the study of electromagnetic waves
in a tokamak. Perhaps fusion research, too, might profit from the
methods developed here.

\section{Retro and prospects}
\label{sec:retrospect}
We have now two systematic approaches to exact solutions of
Maxwell's equations when material properties vary in space.
What was known before this article was written?

The interest in electromagnetically variable media increased dramatically
with the advent of dielectric waveguides, i.e.\ in the seventies
of the previous century. Attempts were made to improve
the fibers using dielectrics with a graded index of refraction
(GRIN), see \cite{Kaw05} for references.
In 1975, Kogelik noted down a Helmholtz equation with a variable
coefficient as in (\ref{hel}) which, according to his belief,
would found a theory of electromagnetic waves in GRIN
media \cite[Sec.2.4]{Kog75}. He remarked the similarity of his equation
with Schr\"odinger's and rewrote some solutions found in textbooks
on quantum mechanics for his purpose. For most reseachers this still
seems to be the state of the art. In 2010, for example,
Lipson, Lipson and Lipson struggle  to derive that Helmholtz equation
from Maxwell's equations \cite[Sec.10.1.2]{Lip10}. They
encounter a term which bears some similarity with the peculiar second term
in (\ref{repa1a}), but cannot handle it and so they discard it
pretending an approximation called \lq\lq weak guidance\rq\rq.
One can find many examples of proclivity for the Helmholtz equation.
Maybe for another confirmation of the slow progress, see Rauh,
Yampolskaya and Yampolskii who reach in 2010 no higher point of view
than Kogelik in 1975. They call \lq\lq Master equation\rq\rq
what others denote as Helmholtz or Schr\"odinger equation. The paper
is valuable nevertheless for some historical references \cite{Rau10}.

In their classic monography, Born and Wolf \cite[Sec.1.6]{Bor68} reproduce
some deliberations probably first thought by Abel\'es \cite{Abe50} in 1950.
Maxwell's equations are written in a cartesian coordinate system
and permittivity as well as permeability are allowed to depend on one
coordinate.
Maxwell's system, which usually couples all components of
the electromagnetic field, is shuffled until there is one equation
for only one component. This equation has the same shape as
(\ref{car3a}) although its physical content is different.
Unfortately this useful equation comes with a second, more complicated
equation which has to be solved at the same time if the
electromagnetic field is to be calculated. In summary,
the theory put forward by Born and Wolf is practically useless.

So Born and Wolf haste to a theory which construes
the continuum as a sequence of small steps.
The graded medium is replaced with a pile of thin layers.
The reflection and the transmission in a single layer
are calculated from Snell's law and Fresnel's formulae.
The results are entered in simple matrices such that the reflections and
the transmissions in the pile can be computed as matrix multiplication.
This is the transfer-matrix method. It is, with
many technical improvements, most popular with practicians. Software packages
that help to construct the matrices and to execute their multiplication
can be found and downloaded in the internet. Have a look, for example,
at \lq\lq Freesnell\rq\rq or \lq\lq RP Coating\rq\rq!

In 2010 Turakulov presented a preprint \cite{Tur10} and in 2011
an article \cite{Tur11} wherein vector potentials were used.
Ordinary differential equations similar to the correct equations
(\ref{car3a}) and (\ref{car3b}) were found
although in a different mathematical context
and a useless scalar potential was introduced \cite{Tur10}
which obscures the calculation of the electromagnetic field.
Turakulov also communicated that the equations (\ref{car3a}) and (\ref{car3b})
can actually be solved for the example (\ref{tur}). Unfortunately
the solutions presented by him were not correct.

The awkwardness of the approaches mentioned so far is caused
by the arbitrary eliminations.
One of the 12 components of the electromagnetic field
and one of 8 Maxwellian equations is selected whereupon lots of
unsystematic attempts are made to eliminate the other 11 components
from the arbitrarily selected equation. This is sometimes feasible
in the cartesian coordinate system. It is cumbersome in cylindrical
coordinates and becomes a nightmare in spherical coordinates.

Most physical laws
are formulated as partial differential equations between between
vector fields and scalars. Thus we have several independent
variables, denoted by physicists as {\it coordinates\/}, and several
dependent ones, denoted by physicists as {\it fields\/}.
Given are partial differential equations where all these variables
are mingled. Wanted are ordinary differential equations each with one
dependent and one independent variable only. To reach the wanted end,
we must perform two separations: a separation of the dependent variables
and a separation of the independent ones. These two kinds of
separation must be kept separate. The fields must be uncoupled
without reference to special coordinates.
This is the idea not comprehended in previous work.

What one must do is this: In a problem with vector fields one must
decompose them into their longitudinal and their transverse parts.
The former can be represented by a scalar potential, the latter by
two vector potentials, the one being a simple vector potential, the
other a vector potential's vector potential. Both vector potentials
must consist of a predetermined vector field times an amplitude which
describes the dynamics. Otherwise there is no chance to arrive at
a differential equation for one scalar quantity only.

This is the recipe that works in all vector-field theories,
e.g.\ in the theory of sound, hydrodynamics, elasticity
and electrodynamics \cite{Bro85}. It was applied in \cite{Bro10}
for a general method solving Maxwell's equations when the material
constants are constant. Finally it was applied here, namely in
(\ref{sol1}-\ref{sol2}) and in (\ref{solra}-\ref{solrb}).

The solutions of the equations (\ref{rep2a}) and (\ref{rep2b})
in the Sections \ref{sec:one} appear all to be new. Seemingly nobody has
noticed that the theory of dielectric mirrors is based on the Mathieu equation
and its generalizations. Calculations up to now were done
using the transfer-matrix method described above; consider the paper
by Fink, Winn, Fan, Michel, Joannopoulos and Thomas
\cite{Fin98} as an anchor of references.
Appearently there were not the least precursors for the solutions
of the spherical problem in Sections \ref{sec:central} ff.\
nor were there harbingers of the two-dimensional theorem of representation
in Section \ref{sec:alternative}, especially of the solution of the
cylindrical problem in Section \ref{sec:axis}.

The broad scope of this article is at the same time its weakness. The
applications were here only indicated, but must be elaborated in detail
to be useful for the design of optical instruments. For example,
in the theory of dielectric mirrors, the permittivity must be
formulated as a Fourier series
\begin{equation}
\epsilon=\sum_{n=-N}^{N} \epsilon_n e^{ink_0z}                \label{mirr}
\end{equation}
with positive wavenumber $k_0$ and
coefficients $\epsilon_n,\ n=0,\pm 1,\pm 2,\ldots\,\pm N$
which model the real medium. Solving the respective equations
(\ref{rep2a}) and (\ref{rep2b}) costs some work, but it can be done
as precisely as wanted. The reward are exact solutions of Maxwell's
equations. These solutions should be better than those obtained using
the transfer-matrix method because gradual transitions between the
layers in the mirror can be taken into account.

O.Gonzales' hunt for references is gratefully acknowledged.

\end{document}